\documentclass[11pt]{article}
\usepackage{amssymb}
\usepackage{amsmath}
\usepackage{amsthm}
\usepackage{caption}
\usepackage{float}
\usepackage{graphicx,color}

\def\proof{{\bf Proof.}}

\def\openone{\leavevmode\hbox{\small1\kern-3.8pt\normalsize1}}
\def\tr{{\rm Tr}}
\def\id{{\rm id}}
\def\ch{{\cal H}}

\def\cb{{\cal B}}
\def\cl{{\cal L}}
\def\cn{{\cal N}}

\def\ci{{\cal I}}

\def\cd{{\cal D}}
\def\cv{{\cal V}}
\def\ck{{\cal K}}
\def\cv{{\cal V}}
\def\cp{{\cal P}}
\def\calr{{\cal R}}
\def\cs{{\cal S}}
\def\ct{{\cal T}}
\def\cu{{\cal U}}
\def\cz{{\cal Z}}

\newcommand{\be}{\begin{equation}}
\newcommand{\ee}{\end{equation}}
\newcommand{\bea}{\begin{eqnarray}}
\newcommand{\eea}{\end{eqnarray}}
\newcommand{\bean}{\begin{eqnarray*}}
\newcommand{\eean}{\end{eqnarray*}}

\def\QQ{\mathbb{Q}}
\def\NN{\mathbb{N}}
\def\HH{\mathbb{H}}
\def\KK{\mathbb{K}}
\def\CC{\mathbb{C}}
\def\DD{\mathbb{D}}
\def\LL{\mathbb{L}}
\def\PP{\mathbb{P}}
\def\UU{\mathbb{U}}
\def\WW{\mathbb{W}}

\def\eff{\text{eff}}
\def\Red{\text{Red}}
\def\omef{\omega_{\eff}}

\def\st{\text{st}}
\def\proof{\textbf{Proof}\quad}

\newtheorem{proposition}{Proposition}

\theoremstyle{definition}
\newtheorem{definition}{Definition}

\newtheorem{remark}{Remark}

\begin{document}
\title{Complete positivity and thermodynamics in a driven open quantum system}
\author{G. Argentieri$^{1}$, F. Benatti$^{2,3}$, R. Floreanini$^3$ and M. Pezzutto$^4$\\[3mm]
\small\it $^1$Department of Physics and Astronomy, University of Exeter,\\
\small\it Stocker Road, Exeter EX4 4QL, UK\\ 
\small\it $^2$Dipartimento di Fisica, Universit\`a di Trieste, I-34151 Trieste, Italy \\
\small\it $^3$Istituto Nazionale di Fisica Nucleare, Sezione di Trieste, I-34151 Trieste, Italy\\[1mm]
\small\it$^4$Physics of Information Group, Instituto de Telecomunica\c{c}\~{o}es \\ 
\small\it and Instituto Superior T\'{e}cnico, Universidade de Lisboa, P-1049-001  
Lisbon, Portugal
}

\date{}

\maketitle

\begin{abstract}
While it is well known that complete positivity guarantees the fulfilment of the second law of thermodynamics, its possible violations have never been proposed as a check of the complete positivity of a given open quantum dynamics.
We hereby consider an open quantum micro-circuit, effectively describable as a two-level open quantum system, whose asymptotic current might be experimentally accessible. This latter could indeed be used to discriminate between its possible non-completely positive Redfield dynamics and a completely positive one obtained by standard weak-coupling limit techniques, at the same time verifying the fate of the second law of thermodynamics in such a context.  
\end{abstract}

\section{Introduction}\label{intro}
\label{sec1}

The so-called open quantum system paradigm is a remarkably successful way to cope with quantum systems weakly interacting with their environments; it has been applied to atomic and molecular physics, quantum optics, quantum chemistry and
condensed matter physics~\cite{alickilendi07,petruccione,weiss,zoller}.

In most physical applications there are no initial statistical correlations between system and environment and their interactions are weak; in such cases, a reduced dynamics of semigroup type for the open quantum system alone can be derived by tracing away the environment degrees of freedom, through the so-called weak-coupling limit techniques \cite{alickilendi07}. The corresponding time-evolution is irreversible and characterized by dissipation and noise; furthermore, it does not only preserve the positivity of the time-evolving states of the system, but is also completely positive~\cite{GKS76,lindblad76,davies76}.
Positivity preservation ensures that the eigenvalues of the density matrices describing the states of the open system remain positive in the course of time so that they can be interpreted as probabilities. Instead, complete positivity guarantees the positivity at all times of any entangled state of compound systems consisting of the open quantum system of interest and any dynamically inert finite level system to which the former may happen to be statistically coupled.

Notice that, when initially there are no correlations between system and environment, the reduced dynamics that results from eliminating the environment degrees of freedom automatically consists of completely positive maps; only, the latter do not in general compose as a semigroup. In order to obtain a forward-in-time composition law, a desirable physical property in a weak coupling regime, one usually performs various kinds of Markovian approximations that often lead to loss of complete positivity and even of positivity \cite{dumcke78}.

The justification for asking that the semigroup dynamics be completely positive because of a general and uncontrollable correlations between the open system and arbitrary inert ancillas, 
is often criticised in the literature as an abstract mathematical artifact~\cite{Pechukas94,Sudarshan05,Rodriguez08,Mc13}.
Indeed, because of complete positivity, the generator of the reduced dynamics assumes the so-called Lindblad form to which there correspond specific physical constraints: typically, a hierarchy among the decay times of the various entries of the density matrices that describe the state of the open quantum system~\cite{alickilendi07,benatti02,benatti05}.
Without complete positivity, less constrained dynamics emerge that render easier the occurrence of so-called ``quantum miracles", like, for instance, the beating of classical bounds in the efficiencies of energy transport processes.

So far, the completely positive character of a dissipative dynamics has only been verified by checking, in very few cases, the above mentioned hierarchy, in general a difficult experimental task.
Instead, in the following, we offer a different strategy based upon the thermodynamic behaviour of open quantum systems. Indeed, complete positivity was soon recognised to imply the positivity of the internal entropy production as required by the second law of thermodynamics~\cite{spohnlebowitz78,spohn78,alicki79}; although there is no proof that complete positivity is also necessary to avoid conflicts with thermodynamics, nevertheless one expects that, in absence of complete positivity, the second law of thermodynamics might indeed be violated. However, no instances either theoretical or experimental have so far been investigated in this respect. In the following, we propose a concrete physical context where to study a possible conflict between a non-completely positive dynamics and the second law of thermodynamics \footnote{A preliminary investigation on these topics has been reported in \cite{argentieri14}.}.

We shall focus upon a model consisting of three electrons moving in
a micro-circuit consisting of a three site loop, under the action of a periodical driving and of a weak coupling to
a thermal bath $\cb$ modelled as a collection of free harmonic oscillators in equilibrium at temperature $T$.
This system, effectively describable as an open $2$-level system, has been introduced to study the dependence on the driving of the asymptotic current that sets in because of the thermal environment~\cite{tosatti}.
In the following, we shall study its behaviour from a thermodynamic point of view and show that the non-completely positive Redfield dynamics considered in~\cite{tosatti} to model the behaviour of the system, violates the second law of thermodynamics.
On the other hand, by adapting standard weak-coupling techniques~\cite{davies76,alickilendi07,dumcke78}, originally developed for time-independent system-environment Hamiltonians, to the present driven system, we derive a completely positive reduced dynamics with explicitly time-dependent generator which gives rise to temporal patterns of the current supported by the microcircuit that significantly differ from those presented in~\cite{tosatti}.
The violations of the second law of thermodynamics by the Redfield-type time-evolutions can then be measured by experimentally accessing the current time-behaviour, thus offering both a test of complete positivity and of the fulfilment of the second law of thermodynamics in open quantum system dynamics.

\section{Thermodynamics of open quantum systems}
\label{sec2}

In this section we provide a brief overview of the approach to the thermodynamics of open quantum systems as formulated in~\cite{alicki79}.

A quantum thermodynamical system is taken to be a finite $n$-level quantum system $\cs$ subjected to a periodically driven time-dependent Hamiltonian $H(t)$ accounting for work being performed on $\cs$ in a cycle. On the other hand, heat flows into and out of $\cs$ because of weak interactions with an environment $\cb$ whose effects are supposed to be effectively described by a master equation with explicitly time-dependent generator $\LL_t$,
\be
\label{eq:thermme}
\frac{d \varrho(t)}{dt} = \LL_t[\varrho(t)]=-i\,[H(t),\varrho(t)] + \DD_t[\varrho(t)] \ ,
\ee
where $\varrho(t)$ is the density matrix describing the state of $\cs$ and $\hbar$ has been set equal to $1$.

The dissipation and noise contributed by the heat bath $\cb$ are accounted for by the time-dependent Lindblad-like contribution
\be
\label{offdiagdiss}
\DD_t[\varrho(t)]=\sum_{j,k} K_{jk}(t)\Big(V_j\,\varrho(t)\,V^\dag _k- \frac{1}{2} \Big\{ V^\dag_k\,V_j\,,\, \varrho(t)\Big\} \Big)\ ,
\ee
with suitable $n\times n$ matrices $V_i$.
The master equation gives rise to a  two-parameter semigroup of trace-preserving maps
$$
\gamma_{t,s}=\ct{\rm e}^{\int_s^t{\rm d}u\,\LL_u}\ ,\quad \gamma_{t,s}\circ\gamma_{s,t_0}=
\gamma_{t,t_0}\ ,\quad \tr\big(\gamma_{t,s}\varrho\big)=\tr\varrho\ ,\qquad t\geq s\geq 0\ ,
$$
where $\ct$ denotes time-ordering.
The complete positivity of such maps is guaranteed by the Kossakowski matrix, $K_t=[K_{jk}(t)]$, being positive-definite for all $t\geq 0$~\cite{alickilendi07}.

Equipped with the formalism of open quantum systems, one can formulate the quantum versions of the laws of classical thermodynamics as follows.

\noindent
$\bullet$\quad
Zero-th law of thermodynamics: it regards the fact that systems initially at different temperatures eventually reach thermal equilibrium.
In the usual open quantum system setting, namely when there is no explicit time-dependence in the generator of~\eqref{eq:thermme}, $\LL_t=\LL$ and $H(t)=H$,
the zero-th law corresponds to the system $\cs$ reaching the thermal equilibrium state
$$
\varrho_\beta=\frac{{\rm e}^{-\beta H}}{\tr\big( {\rm e}^{-\beta H}\big)}\ ,\quad \beta=\frac{1}{\kappa T}\ ,\quad \LL[\varrho_\beta]=0\ ,
$$
at the temperature $T$ of its heat bath environment, $\kappa$ being the Boltzmann constant.
In the present thermodynamic setting with an explicitly time-dependent generator $\LL_t$, the zero-th law amounts to the request that the instantaneous Gibbs state of $\cs$ at time $t$ (at the reservoir temperature), $\varrho_\beta(t)$, be a steady state with respect to
the reduced dynamics:
\be
\label{eq:gibbst}
\varrho_\beta(t)=\frac{{\rm e}^{-\beta H(t)}}{\tr\big( {\rm e}^{-\beta H(t)}\big)}, \qquad \LL_t[\varrho_\beta(t)]= 0 \ .
\ee
This request is physically justified when the system-bath coupling is weak and the external driving by $H(t)$ very slow in comparison with the dissipative time-scale.
\medskip

\noindent
$\bullet$\quad First law of thermodynamics: it concerns the relation between the rate of work done by the system, $\displaystyle \frac{{\rm d}W_t}{{\rm d}t}$, and the rate of heat absorbed from the bath, $\displaystyle \frac{{\rm d}Q_t}{{\rm d}t}$, with the internal energy variation (energy balance):
$$
\frac{{\rm d}E_t}{{\rm d}t}=\frac{{\rm d}Q_t}{{\rm d}t}-\frac{{\rm d}W_t}{{\rm d}t}\ ,
$$
where $\displaystyle E_t = \tr\big(\varrho(t) H(t) \big)$.
The work per unit time done by the open quantum system
due to the time-dependence of the system Hamiltonian is
\be
\frac{{\rm d}W_t}{{\rm d}t}= -\tr \bigg( \varrho(t)\frac{{\rm d}\,H(t)}{{\rm d}t} \bigg)\ .
\ee
Then, the heat exchanged per unit time by the system amounts to
\be
\label{eq:dqdt}
\frac{{\rm d}Q_t}{{\rm d}t} =\frac{{\rm d}(E_t+W_t)}{{\rm d}t}=\tr\bigg(\,H(t)\,\frac{{\rm d}\varrho(t)}{{\rm d}t}\bigg)\ .
\ee
By means of the master equation~\eqref{eq:thermme}, this expression can be recast as
\be
\label{eq:dqdt1}
\frac{{\rm d}Q_t}{{\rm d}t}=\,\tr\bigg(H(t)\,\LL_t[\varrho(t)]\bigg)=\,\tr\bigg(H(t)\,\DD_t[\varrho(t)]\bigg)\ .
\ee

\noindent
$\bullet$\quad Second law of thermodynamics: it regards the variation of the internal entropy of a system, namely the entropy variation which is not due to heat exchanges between the system and the environment and thus not of the form $\displaystyle\frac{\delta Q}{T}$.
The second law asserts that the internal entropy cannot decrease in time.

In order to cast this request in a mathematical form, firstly one introduces the total entropy of the system as given by the von Neumann entropy of its state $\varrho(t)$,
\be
\label{vNe}
S(\varrho(t)) =  -\kappa \tr\Big( \varrho(t) \log\varrho(t) \Big) \ .
\ee
Then, one considers that, as outlined above, its variation in time,
\be
\label{aident}
\displaystyle\frac{{\rm d}S(\varrho(t))}{{\rm d}t}=\frac{{\rm d}S_{\text{int}}(t)}{{\rm d}t}+\frac{{\rm d}S_{\text{ext}}(t)}{{\rm d}t}\ ,
\ee
is due to two terms: an external entropy rate related to the heat exchange with the environment,
\be
\label{eq:entflow}
\frac{{\rm d}S_{\text{ext}}(t)}{{\rm d}t}=\frac{1}{T} \frac{{\rm d} Q_t}{{\rm d}t} =
\kappa \beta\,\tr\Big(H(t) \,\LL_t[\varrho(t)] \Big)\ ,
\ee
and an internal entropy rate $\displaystyle \frac{{\rm d}S_{\text{int}}(t)}{{\rm d}t}$ that we shall denote by $\sigma(\varrho(t))$ for sake of simplicity.

\begin{definition}
\label{def1}
The internal entropy production in a quantum thermodynamic system evolving in time according to the master equation~\eqref{eq:thermme}
is given by
\be
\label{eq:sigmaorig}
\sigma(\varrho(t))=\frac{{\rm d}S_{\text{int}}(t)}{{\rm d}t}=\frac{{\rm d}S(\varrho(t))}{{\rm d}t} - \frac{{\rm d}S_{\text{ext}}(t)}{{\rm d}t}=
- \kappa\tr \bigg( \LL_t[\varrho(t)] \Big(\log\varrho(t) + \beta H(t) \Big) \bigg) \ .
\ee
The second law of thermodynamics amounts to the request that $\sigma(\varrho(t))\geq 0$ for all $t\geq 0$.
\end{definition}
\medskip

The positivity of the internal entropy production is guaranteed by the complete positivity of the maps generated by the master equation~\eqref{eq:thermme}.
This can be seen as follows: using the definition~\eqref{eq:gibbst} of the instantaneous Gibbs state, one has
$$
\tr\bigg(\LL_t[\varrho(t)]\,\beta H(t)\bigg) =-\tr\bigg(\LL_t[\varrho(t)]\,\log\varrho_\beta(t)\bigg)\ ;
$$
then, one can rewrite
\be
\label{eq:entprod}
\sigma(\varrho(t)) = - \kappa\tr\bigg( \LL_t[\varrho(t)]\Big( \log\varrho(t) - \log\varrho_\beta(t) \Big) \bigg) \ .
\ee
The sign of $\sigma(\varrho(t))$ is accessed through  the relative entropy of two density matrices $\varrho_{1,2}$,
\be
S(\varrho_1 \vert \varrho_2) := \tr\bigg( \varrho_1 \Big(\log\varrho_1 - \log\varrho_2 \Big)\bigg)\ ,
\ee
for it monotonically decreases under completely positive, trace-preserving maps~\cite{wehrl78}.

Indeed, for each fixed $t\geq 0$, one can use the generator $\LL_t$ of the master equation~\eqref{eq:thermme} and construct a semigroup of
maps $\Lambda_s=\exp{(s\LL_t)}$, $s\geq 0$. Due to the Linblad form~\eqref{offdiagdiss} of  $\LL_t$ the
maps $\Lambda_s$ are completely positive and trace-preserving; moreover, from $\LL_t[\varrho_\beta(t)]=0$ it follows that $\Lambda_s[\varrho_\beta(t)]=\varrho_\beta(t)$ for all $s\geq 0$. Then,
\bean
S(\Lambda_{s+\delta}[\varrho(t)] \vert\varrho_\beta(t))&=&S(\Lambda_{s+\delta}[\varrho(t)] \vert\Lambda_{s+\delta}[\varrho_\beta(t)])=S(\Lambda_\delta\circ\Lambda_s[\varrho(t)] \vert\Lambda_\delta\circ\Lambda_s[\varrho_\beta(t)])\\
&\leq& S(\Lambda_s[\varrho(t)]\vert \Lambda_{s}[\varrho_\beta(t)])=S(\Lambda_s[\varrho(t)]\vert\varrho_\beta(t))\ .
\eean
By taking the derivative of $-S(\Lambda_{s}[\varrho(t)] \vert\varrho_\beta(t))$ with respect to $s$ at $s=0$, it follows that
\be
\label{posentprod}
0\leq -\kappa\,\frac{{\rm d}S(\Lambda_{s}[\varrho(t)] \vert\varrho_\beta(t)])}{{\rm d}s}\left\vert_{s=0}\right.=-\kappa \tr\bigg(\LL_t[\varrho(t)] \Big(\log{\varrho(t)} - \log{\varrho_\beta(t)} \Big)\bigg)=\sigma(\varrho(t))\ .
\ee

For the interpretation of the term $-\tr\varrho(t)\log\varrho_\beta(t)$ in the relative entropy as a heat flow, it is crucial that
$\varrho_\beta(t)$ be a Gibbs state~\eqref{eq:gibbst} such that $\LL_t[\varrho_\beta(t)]=0$.
In many cases of physical interest, however, the steady states $\varrho_\st(t)$, $\LL_t[\varrho_\st(t)]=0$, need not be Gibbs states.
In these cases, the notion of entropy production can be extended following the argument of~\cite{spohn78}.

\begin{definition}
\label{def2}
In the case of a quantum thermodynamical system undergoing a dissipative dynamics with a stationary state $\varrho_\st(t)$ which is not a Gibbs thermal
state, the expression of the internal entropy production in Definition~\ref{def1} is generalized to
\be
\label{eq:sigmadef}
\sigma(\varrho(t)) := -\kappa\, \frac{{\rm d}}{{\rm d}t} S(\varrho(t) \vert \varrho_\st(t))=-\kappa\tr\bigg(\LL_t[\varrho(t)]
\big(\log\varrho(t)-\log\varrho_\st(t)\big)\bigg)\ ,
\ee
with $\sigma(\varrho(t))\geq 0$ for all $t\geq0$ expressing the second law of thermodynamics.
\end{definition}

That the expression~\eqref{eq:sigmadef} can be interpreted as a {\it bona fide} internal entropy production is based upon
its vanishing when the dynamics is unitary and its convexity as required by thermodynamical stability~\cite{spohn78}.

\section{Open quantum micro-circuit}
\label{sec3}

In this section, we study the second law of thermodynamics in the following solid state model introduced in~\cite{tosatti}: a triangular loop micro-circuit comprising three equal, single orbital quantum dots with three electrons that can jump from one to the other under the action of a periodic modulation
of the transmission amplitudes and gate voltages.
The purpose of~\cite{tosatti} was the study of the asymptotic current which sets in when the micro-circuit $\cs$ is placed in weak interaction with a thermal bath $\cb$ consisting of non-interacting harmonic oscillators.
We briefly present the main characteristics of the model indicating how it can be reduced to the study of an open two-level system.

Let $a,b,c$ label the three quantum dots; in absence of thermal bath, their dynamics is generated by a Hamiltonian that, in the standard basis
$\vert a\rangle$, $\vert b\rangle$ and $\vert c\rangle$ is chosen of the form
$$
H_0(t)=\begin{pmatrix}
\varepsilon_a(t)&-\gamma_0&-\gamma_0\cr
-\gamma_0&\varepsilon_b(t)&-\gamma_0\cr
-\gamma_0&-\gamma_0&\varepsilon_c(t)
\end{pmatrix}
$$
where $\gamma_0>0$ is the hopping amplitude from one dot to the other and $\varepsilon_{a,b,c}(t)$ are the
following external biases:
$$
\varepsilon_a(t)=-\Delta\cos(\Omega t)\ ,\,
\varepsilon_b(t)=-\Delta\cos\Big(\Omega t-2\pi/3\Big)\ ,\ \varepsilon_c(t)=-\Delta\cos\Big(\Omega t+2\pi/3\Big)
$$
with $\Delta\geq 0$ such that $\Delta\ll \gamma_0$ and $\Omega$ the external driving frequency.
The unperturbed Hamiltonian $H_0$ obtained from $H_0(t)$ by setting the biases equal to $0$ has a ground state
$\displaystyle \vert 0\rangle = \frac{\vert a\rangle+\vert b\rangle+\vert c\rangle}{3}$
with energy $-2\gamma_0$ and the following two degenerate orthogonal excited states with energy $\gamma_0$
$$
\vert x\rangle=\frac{\vert b\rangle-\vert c\rangle}{\sqrt{2}}\ ,\quad \vert y\rangle=\frac{2\vert a\rangle-\vert b\rangle-\vert c\rangle}{\sqrt{6}}\ .
$$
In the hypothesis of neglecting electron-electron correlations, when three electrons
move over the micro-circuit, two of them are expected to occupy the ground state, while the state of the third one belongs to the
orthogonal subspace.

As the biases are considered as small perturbations, one may neglect transitions off the ground state and thus restrict to considering the third electron described by the two-dimensional subspace spanned by $\vert x\rangle$ and $\vert y\rangle$.
The restriction of $H_0(t)$ to such a subspace yields a pseudo-spin Hamiltonian
\be
\label{eq:Horiginal}
H(t) = \frac{\Delta}{2} \big(\sigma_3 \cos{\Omega t} + \sigma_1 \sin{\Omega t} \big) \ ,
\end{equation}
where $\sigma_{1,2,3}$ are the standard Pauli matrices with respect to the basis $\vert x\rangle$ and $\vert y\rangle$.

The description of the micro-circuit can thus be effectively reduced to that of a two-level system with time-dependent Hamiltonian.
Within this setting, the electronic current at time $t$ sustained by the micro-circuit is measured by the mean value of $\sigma_2$ with respect to the time-evolving two-level system density matrix $\hat\varrho(t)$:
\be
\label{eq:current1}
I(t) =I_0\tr_{\cs}(\hat\varrho(t)\, \sigma_2) \ ,\ I_0=\frac{e\gamma_0}{\sqrt{3}}\ ,
\ee
where $e$ is the electron charge.
Indeed, the current is induced by the hoppings of the electrons between adjacent dots and thus it is described by the self-adjoint matrix
$$
\hat{I}=ie\gamma_0\Big(\vert a\rangle\langle b\vert-\vert b\rangle\langle a\vert\Big)\ .
$$
By rewriting $\hat{I}$ in the orthonormal basis $\vert 0\rangle$, $\vert x\rangle$, $\vert y\rangle$, and neglecting transition terms connecting the unperturbed ground state  $\vert 0\rangle$,
and the excited states
$\vert x\rangle$ and $\vert y\rangle$, one finds
$$
\hat{I}=\frac{ie\gamma_0}{\sqrt{3}}\Big(\vert y\rangle\langle x\vert-\vert x\rangle\langle y\vert\Big)=\frac{e\gamma_0}{\sqrt{3}}\sigma_2 \ .
$$

The effects of non-negligible interactions of the micro-circuit with the environment that surrounds it will be described by coupling the effective degrees of freedom to a heat bath $\cb$ consisting of independent harmonic oscillators at temperature $T$. Namely,
one considers a total Hamiltonian of the system $\cs+\cb$ of the form $H_{\cs+\cb}(t)=H(t) + H_\cb + H_{\cs\cb}$, where
$$
H_\cb= \sum_{\xi=1,3} \sum_n \bigg(\frac{p^2_{\xi,n}}{2m} + \frac{m \omega_n^2 q^2_{\xi,n}}{2} \bigg)
$$
is the bath Hamiltonian with $q_{\xi,n}$ and $p_{\xi,n}$,  $\xi=1,3$, position and momentum operators of the bath oscillators.
These are in turn coupled to the micro-circuit degrees of freedom by a spin-Bose interaction Hamiltonian:
$$
\lambda\,H_{\cs\cb} = \lambda\,\sum_{\xi=1,3} \sum_n \lambda_n\sqrt{2m\omega_n} \, \sigma_{\xi}\otimes q_{\xi,n}  \ ,
$$
where $\lambda$ is a dimensionless coupling, while the constants $\lambda_n$ are suitable energies associated with the bath spectral density.
\medskip

\begin{remark}
\label{rem3}
Here it is important to notice that the system-bath coupling is taken to be homogeneous, {\it i.e.} the coupling constants $\lambda_n$ do not depend on the index $\xi$  of $\sigma_\xi$.
\end{remark}
\medskip

\noindent
The states $\hat{\varrho}_{\cs\cb}$ of the compound system $\cs+\cb$ will evolve in time according to the explicitly time-dependent Liouville-von Neumann equation
\be
\label{Hamdyn}
\frac{{\rm d}\hat{\varrho}_{\cs\cb}(t)}{{\rm d}t}=-i\,\left[H(t)+H_\cb\,+\,\lambda\,H_{\cs\cb}\,,\,\hat{\varrho}_{\cs\cb}(t)\right]\ .
\ee

Notice that the system $\cs$ Hamiltonian can be recast in the form
$$
H(t)=\frac{\Delta}{2}R(t)\sigma_3\,R^\dag(t)\ ,\quad R(t)={\rm e}^{-i\Omega t\sigma_2/2}\ .
$$
Then, by going to a rotating frame by means of the unitary matrix $R(t)$,
the time dependence in~\eqref{Hamdyn} can be moved from $H(t)$ to the interaction term $H_{\cs\cb}$.
Indeed, by setting
\be
\label{rotframe}
\varrho_{\cs\cb}(t)=R^\dag(t)\,\hat{\varrho}_{\cs\cb}(t)\,R(t)\ ,
\ee
its time-evolution equation reads:
\be
\label{ME1a}
\frac{{\rm d}\varrho_{\cs\cb}(t)}{{\rm d}t}=-i\,\left[H_\eff+\,H_\cb\,+\,\lambda\,H_{\cs\cb}(t)\,,\,\varrho_{\cs\cb}(t)\right]
\ee
with a new time-independent  system Hamiltonian
\be
\label{Heff}
H_\eff=\frac{\Delta\sigma_3-\Omega\sigma_2}{2}
\ee
and explicitly time-dependent interaction term
\be
H_{\cs\cb}(t)=\sum_{\xi=1,3} \sum_n \sqrt{2m\omega_n} \, \sigma_{\xi}(t)\otimes q_{\xi,n}\ ,\quad
\sigma_{\xi}(t)=R^\dag(t)\sigma_\xi\,R(t)\ .
\label{ME1b}
\ee

\begin{remark}
\label{rem4}
In the standard open quantum system approach, with no explicitly time-dependent global Hamiltonian, an equation like~\eqref{ME1a} is the starting point
for the applications of the so-called weak-coupling limit techniques~\cite{alickilendi07} that lead to master equations that involve only the degrees of freedom of the system $\cs$ . These techniques are based on the assumption that the coupling constant is small, $\lambda\ll 1$, and that the initial state of the compound system $\cs+\cb$ has the factorized form $\varrho\otimes\varrho_\cb$, with $\varrho_\cb$ a stationary state of the environment. Then, on a slow time-scale $\tau=t\lambda^2$, the true dynamics of the open quantum system can be well approximated by a semigroup of trace-preserving maps $\gamma_t$ acting on the density matrices of $\cs$.
Only if performed with due care, the approximated reduced dynamics results completely positive; in most cases, too rough manipulations yield maps that do not even preserve the positivity of states. In~\cite{dumcke78}, the source of the problem is identified in the too fast oscillations in time due to the system Hamiltonian dynamics. It is also shown there how a suitable time-average recovers complete positivity.
In Appendix A and B we review these issues adapting the arguments of~\cite{dumcke78} to the explicitly time-dependent equation~\eqref{ME1a}.
\end{remark}
\medskip

An instance of a reduced dynamics which fails to be completely positive is provided by the one adopted in~\cite{tosatti} to model the time-evolution of the three dot micro-circuit.
There, starting with an  initial uncorrelated initial state $\varrho\otimes\varrho_\beta$, where
$\displaystyle\varrho_\beta=\frac{{\rm e}^{-\beta H_\cb}}{\tr{\big(\rm e}^{-\beta H_\cb}\big)}$
is the thermal bath equilibrium state at temperature $T$, a Redfield-type
master equation for the micro-system alone is presented:
\bea
\nonumber
\frac{{\rm d}\varrho(t)}{{\rm d}t}=-i\,\Big[H_\eff, \varrho(t)\Big]\,&-&\,\lambda^2\,\sum_{\xi=1,3} \int_0^{+\infty}
{\rm d}u
\,\bigg\{G(u) \bigg[\sigma_{\xi}\,,\, U_\eff(u)\,\sigma_{\xi}(-u)\,U^\dag_\eff(u)\,\,  \varrho(t) \bigg]	\\
&+& G^*(u) \bigg[ \varrho(t)\, U_\eff(u)\, \sigma_{\xi}(-u)\,U^\dag_\eff(u)\,,\,\sigma_\xi \bigg] \bigg{\}}
\ .
\label{eq:metosatti}
\eea
In the above expression, $\varrho(t)=\tr_\cb\Big(\varrho_{\cs\cb}(t)\Big)$ is the density matrix describing the state of the micro-circuit $\cs$ at time $t$,
\be
\label{aido}
U_\eff(u) = \exp{(-i\,u\,H_{\eff})}
\ee
is the unitary time-evolution due to the system $\cs$ Hamiltonian $H_\eff$, while
the complex function
\be
\label{eq:eqcorr}
G(u)=\int_0^{\infty} {\rm d} \omega \, J(\omega)
\bigg{[} \cos{(\omega u)} \coth{\frac{\omega \beta}{2}} - i \sin{(\omega u)} \bigg{]}
\ee
is constructed with the bath two-point correlation functions
\be
{\rm Tr}_\cb\Big(\rho_\beta q_{\xi_1,n_1}\,{\rm e}^{-iu\,H_\cb}\,q_{\xi_2,n_2}\,{\rm e}^{iu\,H_\cb}\Big)
=\delta_{\xi_1,\xi_2}\delta_{n_1,n_2}\,
\frac{\cos(\omega_{n_1} u)\, \coth{\frac{\omega_{n_1} \beta}{2}} - i \sin(\omega_{n_1} u)}{2m\omega_{n_1}}\ ,
\label{2pointf}
\ee
and the bath spectral density
\be
\label{spdens}
J(\omega) = \sum_n\,\lambda^2_n\, \delta(\omega - \omega_n) \, .
\end{equation}
For an infinite heat bath, $J(\omega)$ is taken to be of Ohmic form with cut-off frequency $\omega_c$:
\be
\label{Ohmic}
J(\omega)=\omega\exp(-\omega/\omega_c)\ .
\ee
\medskip

\begin{remark}
\label{rem5}
Notice that, although the Liouville-von Neumann equation~\eqref{ME1a} depends explicitly on time, the master equation~\eqref{eq:metosatti} resulting from the Markovian approximation does not.
As shown in Appendix A, this is the consequence of the homogeneous coupling between system and bath (see Remark~\ref{rem3}).
\end{remark}
\medskip

Because of the disappearance of the explicit time-dependence, the reduced dynamics is represented by a one-parameter semigroup $\{ {\gamma}_t\}_{t\geq0}$
of trace-preserving maps $ {\gamma}_t$ which, however, are not completely positive.
\medskip

\begin{proposition}
\label{prop1}
The semigroup generated by~\eqref{eq:metosatti} consists of non-completely positive maps.
\end{proposition}

\proof
This is easily checked by setting
\bea
\label{V1}
V_1&=&\int_0^{+\infty}{\rm d}\tau\,G(\tau)\, W_1(\tau)\ ,\quad W_1(\tau)=U_\eff(\tau)\, \sigma_1(-\tau)\, U^\dag_\eff(\tau)\ ,\quad V_2=\sigma_1\\
\label{V3}
V_3&=&\int_0^{+\infty}{\rm d}\tau\, G(\tau)\, W_2(\tau)\ ,\quad W_2(\tau)=U_\eff(\tau)\, \sigma_3(-\tau)\, U^\dag_\eff(\tau)\ ,\quad V_4=\sigma_3\ .
\eea
Then, the right hand side of~\eqref{eq:metosatti} can be recast in the form
\be
\label{Lind1}
\LL^{\Red}[\varrho(t)]=-i\,\Big[H_\eff+\lambda^2H^{\Red}_{LS}\,,\,\varrho(t)\Big] + \DD^{\Red}[\varrho(t)]
\ee
with a Lamb-shift Hamiltonian term
\be
\label{Lind2}
H^{\Red}_{LS} = \frac{1}{2i}\Big(V_2V_1-V_1^{\dag}V^\dag_2+V_4V_3-V^\dag_3V^\dag_4\Big)\ ,
\ee
and a purely dissipative contribution
\be
\label{purediss}
\DD^{\Red}[\varrho(t)]=\sum_{j,k=1}^4 K_{jk}
\bigg{(} V_k \varrho(t) V_j^{\dag} - \frac{1}{2}{\{} V_j^{\dag}V_k, \varrho(t) {\}} \bigg{)}
\ee
with Kossakowski matrix given by $\displaystyle K = [K_{jk}]=
\begin{pmatrix}
\sigma_1& 0\\
0	& \sigma_1
\end{pmatrix}$, $\sigma_1=
\begin{pmatrix}
0&1\\
1&0
\end{pmatrix}$.
Since $K=[K_{jk}]$ is not positive definite then the generated maps is not completely positive~\cite{GKS76}.

\subsection{Failure of positivity preservation by the Redfield dynamics}

As previously mentioned, reduced dynamics of Redfield type fail in general to preserve even the positivity of the time-evolving density matrices: this is indeed the case also for the time-evolution used in \cite{tosatti}.
The appearance of negative eigenvalues in the spectrum of $\varrho(t)$ is typically occurring at short times.
Indeed, given the master equation~\eqref{eq:metosatti} with generator in the form~\eqref{Lind1}, an expansion of the solution 
$\varrho(t)$ to first order in $t$ starting with a pure state $\vert\psi\rangle\langle\psi\vert$ yields:
$$
\varrho(t)\simeq\vert\psi\rangle\langle\psi\vert\,+\,t\,\LL^\Red[\vert\psi\rangle\langle\psi\vert]\ .
$$
By choosing $\vert\phi\rangle\in\CC^2$ such that $\langle\phi\vert\psi\rangle=0$, one finds that only the first term in~\eqref{purediss} contributes to the mean value of $\varrho(t)$ at short times:
\bea
\label{mv1}
\langle\phi\vert\varrho(t)\vert\phi\rangle&\simeq&\, 2\,t\,\Delta_{\psi,\phi}\\
\label{mv2}
\Delta_{\psi,\phi}&=&\mathcal{R} e\Big\{\langle\phi\vert V_1\vert\psi\rangle\langle\psi\vert\sigma_1\vert\phi\rangle+
\langle\phi\vert V_3\vert\psi\rangle\langle\psi\vert\sigma_3\vert\phi\rangle\Big\}\ .
\eea
If an orthonormal basis $\vert\psi\rangle, \vert\phi\rangle$ can be found in $\CC^2$ such that $\Delta_{\psi,\phi}<0$, then, for small times, the projector $\vert\psi\rangle\langle\psi\vert$ transforms into an operator of trace one with one negative eigenvalue, otherwise all mean values~\eqref{mv1} should be non-negative.

Choosing $\displaystyle\vert\psi\rangle=\frac{\vert 0\rangle+\alpha\vert 1\rangle}{\sqrt{1+|\alpha|^2}}$ and $\displaystyle\vert\phi\rangle=\frac{\alpha^*\vert 0\rangle-\vert 1\rangle}{\sqrt{1+|\alpha|^2}}$, where $\alpha\in\CC$ and $\sigma_3\vert0\rangle=\vert0\rangle$, $\sigma_3\vert 1\rangle=-\vert1\rangle$, one gets $\Delta_{\psi,\phi}=\frac{\Delta(\alpha)}{(1+|\alpha|^2)^2}$, where
\bean
\Delta(\alpha)&=&\mathcal{R}e\Big\{
\Big((\alpha^*)^2-1\Big)\Big(\alpha(\langle0\vert V_1\vert 0\rangle-\langle1\vert V_1\vert 1\rangle)+\alpha^2\langle0\vert V_1\vert 1\rangle
-\langle1\vert V_1\vert 0\rangle\Big)\\
&+&2\alpha^*\Big(\alpha(\langle0\vert V_3\vert 0\rangle-\langle1\vert V_3\vert 1\rangle)+\alpha^2\langle0\vert V_3\vert 1\rangle
-\langle1\vert V_3\vert 0\rangle\Big)\Big\}\ .
\eean
In order not to violate the positivity of $\displaystyle{\rm e}^{t\,\LL^\Red}[\vert\psi\rangle\langle\psi\vert]$ at short times, $\Delta(\alpha)$ must be non-negative for all $\alpha$.

Setting
$C=\cos(\Omega\,\tau)$, $S=\sin(\Omega\,\tau)$, $c=\cos(\omega_\eff\,\tau)$ and $s=\sin(\omega_\eff\,\tau)$, where
$\displaystyle \omef=\sqrt{\Omega^2 + \Delta^2}$, an explicit evaluation of the $2\times 2$ matrices $W_{1,3}$ in \eqref{V1} and \eqref{V3} yields
\bean
W_1&=&
\left(
\begin{array}{cc}
 \frac{\Omega  \left(\text{C} \text{s}-\frac{\text{c} \text{S} \Omega }{\omef}\right)}{\omef}-\frac{\text{S} \Delta
   ^2}{\left(\omef\right)^2} & \text{c} \text{C}-i \left(\frac{\text{S} \Delta  \Omega }{\left(\omef\right)^2}+\frac{\Delta 
   \left(\text{C} \text{s}-\frac{\text{c} \text{S} \Omega }{\omef}\right)}{\omef}\right)+\frac{\text{s} \text{S} \Omega }{\omega
   '} \\
 \text{c} \text{C}+i \left(\frac{\text{S} \Delta  \Omega }{\left(\omef\right)^2}+\frac{\Delta  \left(\text{C} \text{s}-\frac{\text{c}
   \text{S} \Omega }{\omef}\right)}{\omef}\right)+\frac{\text{s} \text{S} \Omega }{\omef} & \frac{\text{S} \Delta
   ^2}{\left(\omef\right)^2}-\frac{\Omega  \left(\text{C} \text{s}-\frac{\text{c} \text{S} \Omega }{\omef}\right)}{\omef} \\
\end{array}
\right)\\
W_3&=&
\left(
\begin{array}{cc}
 \frac{\text{C} \Delta ^2}{\left(\omef\right)^2}+\frac{\Omega  \left(\text{s} \text{S}+\frac{\text{c} \text{C} \Omega }{\omef}\right)}{\omef} & \text{c} \text{S}-i \left(\frac{\Delta  \left(\text{s} \text{S}+\frac{\text{c} \text{C} \Omega }{\omef}\right)}{\omef}-\frac{\text{C} \Delta  \Omega }{\left(\omef\right)^2}\right)-\frac{\text{C} \text{s} \Omega }{\omef} \\
 \text{c} \text{S}+i \left(\frac{\Delta  \left(\text{s} \text{S}+\frac{\text{c} \text{C} \Omega }{\omef}\right)}{\omef}-\frac{\text{C} \Delta  \Omega }{\left(\omef\right)^2}\right)-\frac{\text{C} \text{s} \Omega }{\omef} & -\frac{\text{C} \Delta
   ^2}{\left(\omef\right)^2}-\frac{\Omega  \left(\text{s} \text{S}+\frac{\text{c} \text{C} \Omega }{\omef}\right)}{\omef} \\
\end{array}
\right)
\eean
By means of these expressions the matrix $V_{1,3}$ in \eqref{V1} and \eqref{V3} can be numerically computed as well as the behaviour of $\Delta(\alpha)$. The following figure shows the complex $\alpha$'s for which $\Delta(\alpha)<0$ and thus the amount of pure states that are sent out of the Bloch sphere at small times by getting a negative eigenvalue.

\begin{figure}[H]
\centering
\label{neg2}
\includegraphics[scale=0.8]{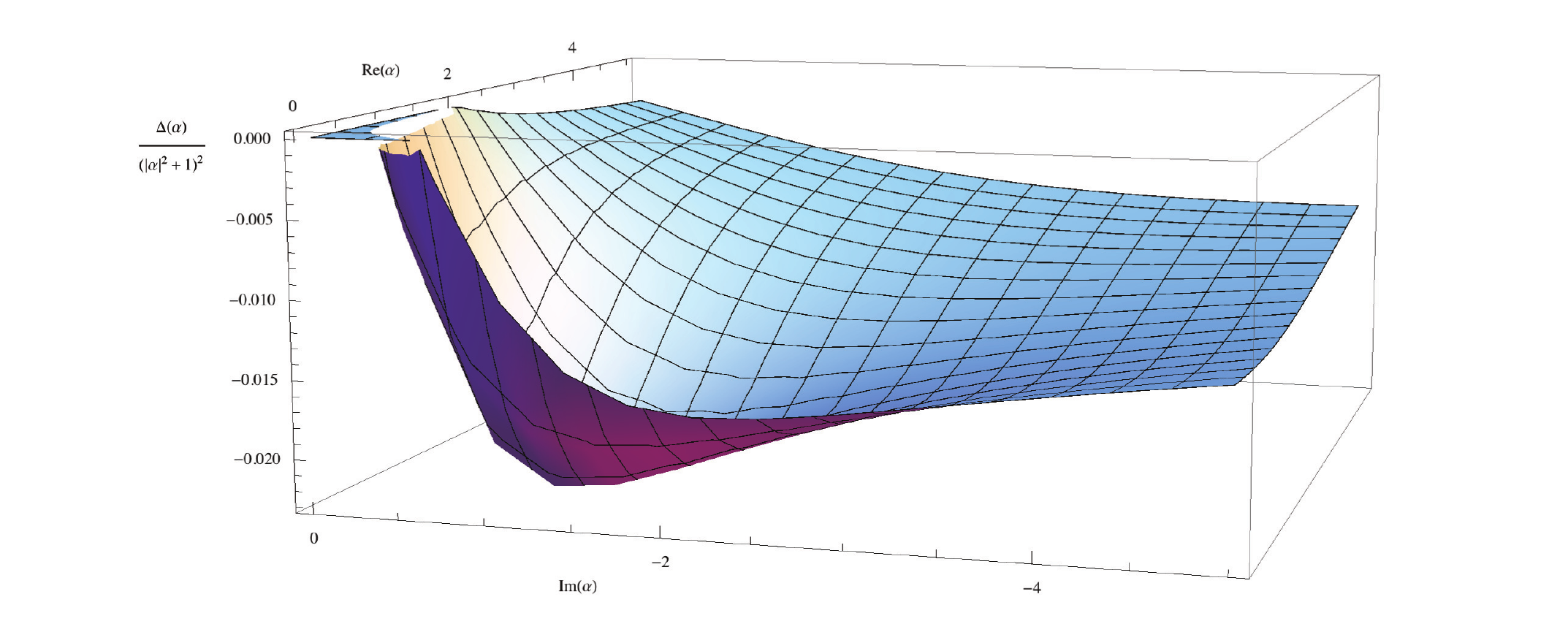}
\caption{\small Presence of negative eigenvalues at short times: $\Delta(\alpha)<0$.}
\end{figure}

\subsection{Completely positive dynamics}

A fully physically consistent time-evolution for the reduced system can be obtained by means of a more careful manipulation through standard methods in open quantum system theory, thus avoiding drawbacks as absence of positivity and of complete positivity. These methods, known as weak-coupling limit, are essentially based upon eliminating too fast oscillations due to the dynamics generated by $H_\eff$ by means of a time-average; however, they are developed for time-independent system-environment Hamiltonians, whereas in the present case we deal with a driven two-level system whence the standard weak-coupling limit techniques must be adapted to the new setting.
This is done in Appendix B, where it is shown in detail how to derive a master equation of the form
\be
\label{MECP}
\frac{{\rm d}\varrho(t)}{{\rm d}t}=\LL[\varrho(t)]=-i\,\Big[H_\eff+\lambda^2H_{LS}\,,\,\varrho(t)\Big]
+\lambda^2\DD[\varrho(t)]
\ee
that generates completely positive maps.

In order to compare the previous master equation with the one in~\eqref{eq:metosatti}, it proves convenient to recast them both in a vectorial form using the so-called Bloch-representation with respect to a new triple of Pauli matrices
\be
\label{eq:newsigma}
\hat{\sigma}_1=\sigma_1\ ,\quad
\hat{\sigma}_2=\frac{\Delta \sigma_2 + \Omega \sigma_3}{\omef}\ , \qquad
\hat{\sigma}_3=\frac{\Delta \sigma_3 - \Omega \sigma_2}{\omef} \ ,
\ee
so that the effective Hamiltonian~\eqref{Heff} becomes
$\displaystyle H_\eff = \frac{\omef}{2} \hat{\sigma}_3$.

One can then expand the time-evolving density matrix as follows,
\be
\label{Bloch}
\varrho(t)=\frac{1}{2}\Big(1+\sum_{i=1}^3r_i(t)\hat{\sigma}_i\Big)\ ,\quad r^2_1(t)+r_2^2(t)+r_3^2(t)\leq 1\ ,
\ee
and identifies it with the real Bloch vector $\vert\varrho(t)\rangle$ with components $(1,r_1(t),r_2(t),r_3(t))$.

Then, the action of the generator $ {\LL}^{\Red}$ in~\eqref{Lind1} can be rewritten  in terms of a $4\times 4$ matrix
$ {\cl}^\Red= {\ch}_\eff\,+\lambda^2\, {\ch}^\Red_{LS}\,+\lambda^2\, {\cd}^\Red$ such that
\be
\label{eq:metind}
\frac{{\rm d}}{{\rm d} t}\begin{pmatrix}1\cr  {r}_1(t)\cr  {r}_2(t)\cr  {r}_3(t)\end{pmatrix}=-2\, {\mathcal{L}}^\Red\,
\begin{pmatrix}1\cr  {r}_1(t)\cr  {r}_2(t)\cr  {r}_3(t)\end{pmatrix}\ ,\quad {\mathcal{L}}^\Red=
\begin{pmatrix}
0		& 0	 		& 0			& 0				\\
 {\cl}_{10} &  {\cl}_{11}	&  {\cl}_{12} 	&  {\cl}_{13}		\\
 {\cl}_{20} &  {\cl}_{21}	&  {\cl}_{22}	&  {\cl}_{23}		\\
 {\cl}_{30}	&  {\cl}_{31}	&  {\cl}_{32}	&  {\cl}_{33}
\end{pmatrix}\ ,
\ee
where the contributions
\be
\label{MECP3}
 {\ch}_\eff=\begin{pmatrix}0&0&0&0\\
0&0&\omef/2&0\\
0&-\omef/2&0&0\\
0&0&0&0
\end{pmatrix} \ ,\quad
 {\ch}^\Red_{LS}=\begin{pmatrix}0&0&0&0\\
0&0& {\ch}_{12}& {\ch}_{13}\\
0&- {\ch}_{12}&0& {\ch}_{23}\\
0&- {\ch}_{13}&- {\ch}_{23}&0
\end{pmatrix}\ ,
\ee
come from the two commutators in~\eqref{Lind1}, while the dissipative term corresponds to
\be
\label{MECP3c}
 {\cd}^\Red=
\begin{pmatrix}
0		& 0	 		& 0			& 0				\\
 {\ck}_{10} &  {\ck}_{11}	&  {\ck}_{12} 	&  {\ck}_{13}		\\
 {\ck}_{20} &  {\ck}_{21}	&  {\ck}_{22}	&  {\ck}_{23}		\\
 {\ck}_{30}	&  {\ck}_{31}	&  {\ck}_{32}	&  {\ck}_{33}
\end{pmatrix}\ .
\ee
The coefficients of these matrices are explicitly given in Appendix A. Instead, in Appendix B, the generator of the master equation~\eqref{MECP} is proved to correspond to the $4\times 4$ matrix
$\cl=\ch_\eff+\lambda^2\, {\ch}_{LS}+\lambda^2\, {\cd}$, where $\ch_\eff$ is as in~\eqref{MECP3},
while
\be
\label{MSEP4}
 {\ch}_{LS}=\begin{pmatrix}0&0&0&0\\
0&0& {\ch}_{12}&0\\
0&- {\ch}_{12}&0&0\\
0&0&0&0
\end{pmatrix}
\ ,\qquad
 {\cd}=\begin{pmatrix}0&0&0&0\\
0& \mathcal{K}_{11}+ \mathcal{K}_{22}&0&0\\
0&0& \mathcal{K}_{11}+ \mathcal{K}_{22}&0\\
 \mathcal{K}_{30}&0&0& \mathcal{K}_{33}
\end{pmatrix}\ .
\ee

\subsection{Stationary states and current}
\label{sec32}

The vectorial form of the master equation in the Bloch representation is particularly suited to finding the stationary states $\varrho_{\st}$ such that $ {\LL}[ {\varrho}_{\st}]=0$; indeed, they correspond to the Bloch vectors $(r_0,r_1,r_2,r_3)$ that solve the linear system
$$
\sum_{\mu=0}^3 {\cl}_{i\mu} {r}_\mu=0\ ,\quad i=1,2,3\ ,\quad r_0=1\ .
$$
The unique stationary state of the master equation~\eqref{MECP} is easily found using~\eqref{MSEP4}: it corresponds to the Bloch vector
\be
\label{eq:state}
 {\varrho}_\st=\frac{1 +  {r}_3^\st \hat{\sigma}_3}{2} \ ,\qquad  {r}_3^\st
= - \frac{ {\ck}_{30}}{ {\ck}_{33}} \ .
\ee
Therefore, in the case of the completely positive reduced dynamics generated by~\eqref{MECP}, all initial states of the open quantum system will asymptotically tend to $ {\varrho}_\st$.

On the contrary, in the case of the non-completely positive Redfield dynamics generated by~\eqref{eq:metosatti}, the stationary state will not in general be unique and the manifold of asymptotic states will depend on the explicit form of the coefficients in the $4\times 4$ matrices $ \mathcal{H}^\Red_{LS}$ and $ \mathcal{D}^\Red$.
In~\cite{tosatti}, based on numerical evidences, it is argued that, with the following choices of coupling constant, $\lambda = 0.005$, bath temperature, \hbox{$T\simeq 0.006$ $\text{K}$},
pumping amplitude, $\Delta\simeq8\,\text{GHz}$, pumping frequency, $\Omega$, and Ohmic cut-off frequency, $\omega_c$,
such that $\omega_c / \Delta= 10^3$, $\Omega/\Delta=2$,
the stationary state in~\eqref{eq:state} can be taken as an approximation of the true stationary state of the Redfield type master equation~\eqref{eq:metosatti}. However, as we shall show below (see Figures~\ref{fig:curr1} and~\ref{fig:curr2}), the 
asymptotic behaviours of the two dynamics give rise to different values for the asymptotic current supported by them.

In the non-rotating frame, the stationary states become time-dependent steady states.
Indeed, the physical reduced dynamics that sends the initial state $\varrho$ into $\hat{\varrho}(t)$ is obtained from the one sending $\varrho$ into
$\varrho(t)$ by going back to the non-rotating representation, where
$\displaystyle\hat{\varrho}(t)=R(t)\,\varrho(t)\,R^\dag(t)$.
However, since $R(t)$ leaves $\sigma_2$ invariant, the currents supported by the micro-circuit in the states $\hat{\varrho}(t)$ and $ \varrho(t)$
are the same:
\be
\label{eq:idc}
I_t = I_0\,\tr\Big(\hat{\varrho}(t)\sigma_2\Big)=I_0\,\tr\Big(\varrho(t)\,\sigma_2\Big)\ .
\ee
In particular, the completely positive time-evolution predicts an asymptotic current supported by the stationary state which, using the explicit
expressions~\eqref{appint12c} and~\eqref{appint13f} of the coefficients $ {\ck}_{30}$ and $ {\ck}_{33}$ computed with the Ohmic spectral density~\eqref{Ohmic}, reads
\be
\label{eq:pst}
I_\st=I_0 \, \frac{\Omega}{\omef} \, \frac{ {\ck}_{30}}{ {\ck}_{33}}\, ,   \qquad
\frac{ {\ck}_{30}}{ {\ck}_{33}} =  \frac{(\omef - \Omega)^2 J_{+} + (\omef + \Omega)^2 J_{-}}
{(\omef - \Omega)^2 c_{+}J_{+} + (\omef  +\Omega)^2 c_{-}J_{-}} \ ,
\ee
where,
$$
J_{\pm} = J(\omef \pm \Omega), \quad c_{\pm} = \coth{\bigg{(} \frac{\beta (\omef \pm \Omega)}{2} \bigg{)}} \ .
$$

\paragraph{Asymptotic and thermal states}
\label{sec321}

In the expressions~\eqref{eq:entprod} and~\eqref{eq:sigmadef} for the internal entropy production there appears either a Gibbs state associated to a heat flux and to the external entropy production, or a stationary (steady) state.

In the thermodynamic analysis of the present model, we have three asymptotic states: the stationary state $ {\varrho}_\st$ in~\eqref{eq:state}, the Gibbs state associated with the micro-circuit Hamiltonian $H_\eff$,
\be
\label{Gibbs1}
 {\varrho}_\beta^\eff=\frac{{\rm e}^{-\beta H_\eff}}{\tr{\rm e}^{-\beta H_\eff}} = \frac{1 - \tanh{(\omef \beta/2)}\hat{\sigma}_3}{2} \ ,
\end{equation}
and the Gibbs state relative to the Hamiltonian $H_\eff$ corrected by the Lamb-shift contribution $ {H}_{LS}^\Red$,
\be
\label{Gibbs2}
 {\varrho}_\beta=\frac{{\rm e}^{-\beta (H_\eff + \lambda^2 H_{LS})}}{\tr{\rm e}^{-\beta (H_\eff + \lambda^2 H_{LS})}} \ .
\ee
In general, $\varrho_\st$ may differ significantly from the two Gibbs states above, their proximity or not being measured by the trace distance in terms of their Bloch expansion coefficients
$$
d(\varrho_a,\varrho_b)=\frac{1}{2}\text{Tr}\sqrt{(\varrho_a-\varrho_b)^2}=
\frac{1}{2}\,\sqrt{(r_1^a-r_1^b)^2 + (r_2^a-r_2^b)^2 + (r_3^a-r_3^b)^2}\ .
$$
It can numerically be checked that the trace-distances between the Gibbs states~\eqref{Gibbs1} and~\eqref{Gibbs2} and the stationary state
$\varrho_\st$ are negligible under the same choice of parameters mentioned above in~\cite{tosatti}.

By the continuity properties of the relative entropy with respect to the trace-norm, it also follows that, under those physical conditions, the Definitions~\ref{def1} and~\ref{def2} of internal entropy production practically coincide.
In the following section we will first deal with the fate of the second law of thermodynamics in relation to the Redfield and the completely positive dynamics without assuming that the stationary state be of Gibbs form; that is we will use the expression~\eqref{eq:sigmadef} for the internal
entropy production. Only when investigating the matter numerically we will fix the above mentioned physical conditions and consider the expression~\eqref{eq:entprod}, taken advantage of the above mentioned proximity of the two definitions of entropy production.

\section{Complete positivity and entropy production}

Although, as discussed above, the stationary state of the micro-circuit reduced dynamics is
explicitly time-dependent, $\hat{\varrho}_\st(t)=R^\dag(t)\, \varrho_\st\,R(t)$, in order to study the internal entropy production, we can consider the time-independent master equations and the stationary states $\varrho_\st$. Indeed, the unitary connection between the physical and the rotated representation yields
\bea
\nonumber
\sigma(\hat{\varrho}(t))&=&- \kappa\, \tr\bigg(\LL_t[\hat{\varrho}(t)]\Big(\log\hat{\varrho}(t) - \log\hat{\varrho}_\st(t)\Big)\bigg)\\
\label{eq:aid}
&=&-\kappa\,
\tr\bigg( {\LL}[\varrho(t)]\Big(\log\varrho(t) - \log\varrho_\st\Big)\bigg)=\sigma(\varrho(t))\ .
\eea
By means of the Bloch parametrization~\eqref{Bloch} of $\varrho(t)$ and of the matrix $\cl$ in~\eqref{eq:metind}, one gets
$$
\LL[\varrho]=-\sum_{i=1}^3\sum_{\mu=0}^3 {\cl}_{i\mu} {r}_\mu(t)\,\hat{\sigma}_i\ ,\qquad  {r}_0=1\ ,
$$
while the spectral representation of $\varrho(t)$ yields
$$
\log\varrho(t)=\frac{1}{2}\left(1+\sum_{j=1}^3\frac{ {r}_j(t)}{ {r}(t)}\hat{\sigma}_j\right)\,\log \bigg( \frac{1+ r(t)}{2} \bigg) \,+
\frac{1}{2} \left(1-\sum_{j=1}^3\frac{ {r}_j(t)}{ r(t)}\hat{\sigma}_j\right)\, \log \bigg( \frac{1- {r}(t)}{2}  \bigg) \ ,
$$
where $ [r(t)]^2=\sum_{j=1}^3  [r_j(t)]^2$.
Then,
\be
\label{eq:sigma_eq}
\sigma(\varrho(t))=\kappa\,\sum_{i=1}^3\sum_{\mu=0}^3\cl_{i\mu} {r}_\mu(t)\left(\frac{ {r}_i(t)}{r(t)} \log\frac{1+ r(t)}{1- r(t)}-\frac{ {r}^\st_i}{ {r}^\st} \log\frac{1+ {r}^\st}{1- {r}^\st} \right)
\ ,
\ee
where $ [r^\st]^2=\sum_{j=1}^3 [{r}^\st_j]^2$.

\paragraph{Internal entropy production: numerical analysis}

\begin{figure}[h]
\centering
\includegraphics[width=0.45\textwidth]{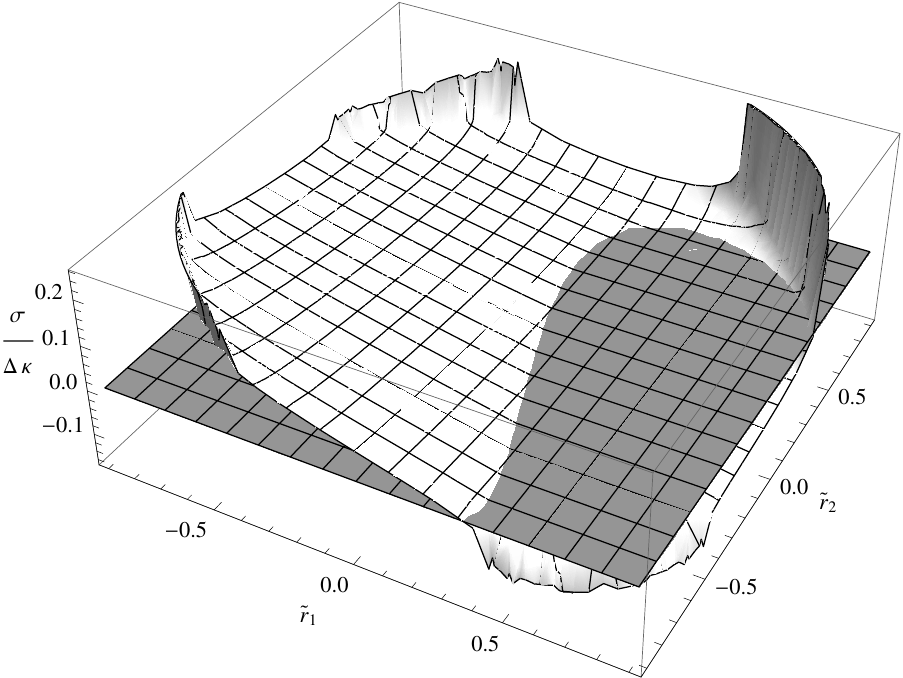}
\caption{\small $\sigma(\varrho)$ as a function of $ {r}_{1,2}$, Redfield dynamics.}
\label{fig:plot3d1}
\end{figure}

We first study the internal entropy production $\sigma(\varrho)$ at $t=0$ as a function of the initial state $\varrho$.
Three dimensional plots can be obtained by computing $\sigma(\varrho)$ in terms of two Bloch vector components:
by setting $ {r}_3=0$ and plotting $\sigma(\varrho)$ as a function of $ {r}_1$ and $ {r}_2$, one sees
that, in the case of the Redfield dynamics generated by $ {\cl}^\Red$, there are regions where the entropy production is negative (see
Fig.~\ref{fig:plot3d1}). None of these violations appear if the reduced
dynamics is completely positive as that generated by $ {\cl}$.

Typically, the physical inconsistencies associated with the appearance of negative probabilities either in the spectrum of a time-evolving density matrix of the open quantum system (lack of positivity) or in the spectrum of an entangled state of the open quantum system and any finite level system (lack of complete positivity), manifest themselves at very short times.
Quite different is the case with the second law of thermodynamics; its violations at time $t=0$ are in fact not a negligible transient effect.
Indeed, a numerical computation of~\eqref{eq:sigma_eq} as a function of time, with physical parameters chosen as in~\cite{tosatti} (see previous section),
shows repeated violations of the second law also in the course of time.
These violations occur for a same percentage of initial states as the violations at $t=0$ and involve also states that initially have $\sigma(\rho)\geq 0$ and thus show no violations of the second law at $t=0$.

\begin{figure}[ht]
\centering
\includegraphics[width=0.50\textwidth]{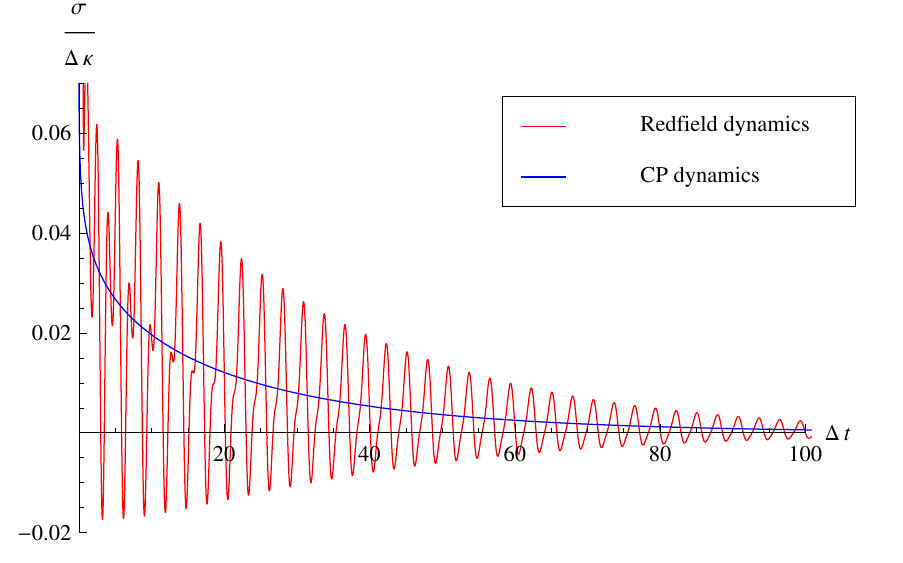}
\caption{\small $\sigma(\varrho(t))$ as a function of time: initial state $\vert\varrho\rangle=(1,0,2r,r)$, $r=-1/\sqrt{5}$.}
\label{fig:sigma_100_2}
\end{figure}

As an example, take as initial state the one with Bloch components $\vert\varrho\rangle=(1,0,2r,r)$, $r=-1/\sqrt{5}$ (cfr.~\eqref{Bloch}): it corresponds to  an eigenstate of $\sigma_3$ and was studied in~\cite{tosatti}. It exhibits an initial $\sigma(\varrho_{t=0})>0$ followed by periodic violations of $\sigma(\varrho(t))\geq0$ (see Figure~\ref{fig:sigma_100_2}).
\vfill\break

\begin{figure}[ht]
\centering
\includegraphics[width=0.50\textwidth]{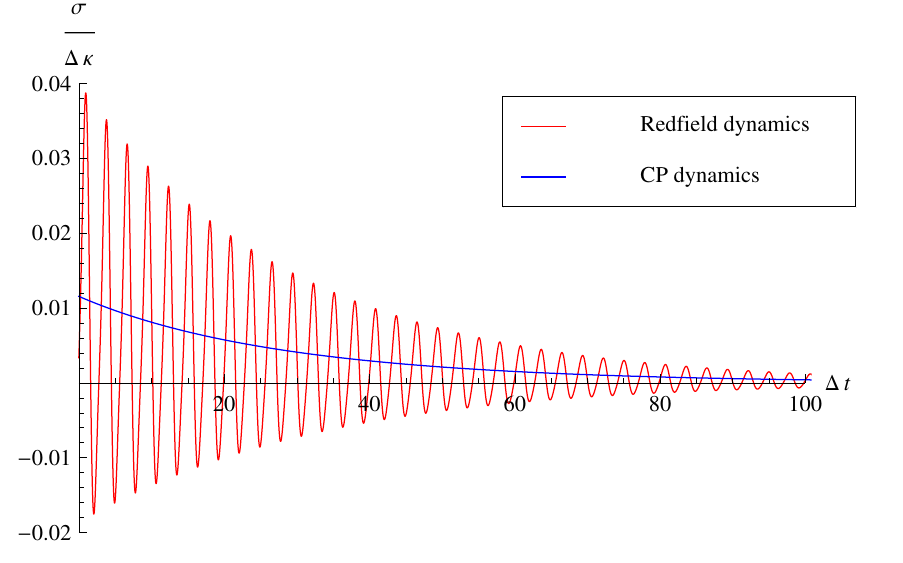}
\caption{\small $\sigma(\varrho(t))$ as a function of time: initial mixed state $\vert\varrho\rangle=(1,0,0.5,-0.4)$.}
\label{fig:sigma_100_3}
\end{figure}

In the case of a mixed state $\vert\varrho\rangle=(1,0,r_2,r_3)$, $r_2=0.5$ and $r_3=-0.4$, that starts with $\sigma(\varrho_{t=0})<0$, again periodic violations of $\sigma(\varrho(t))\geq 0$ appear (see Figure~\ref{fig:sigma_100_3}).

In the graphs of both Figures~\ref{fig:sigma_100_2} and~\ref{fig:sigma_100_3}, the blue line corresponds to the completely positive dynamics generated by $ {\cl}$
that always yields $\sigma(\varrho(t))\geq 0$ in agreement with the theory~\cite{dumcke78}.

\begin{remark}
\label{rem6}
The observed violations of the second law of thermodynamics are not restricted to the specific choice of initial state and physical parameters considered in 
Figure \ref{fig:sigma_100_3} for sake of comparison with the experimental context devised in \cite{tosatti}.
Indeed,  the conflict between the non-complete positivity of the Redfield dynamics and the non-negative internal entropy production
manifests itself across a whole range of parameters, namely for temperatures between $0.0006$ $K$ and $0.06$ $K$ 
and ratios $\Omega / \Delta$ between $0.1$ to $10$.
In particular, violations of the second law of thermodynamics at time $t=0$ always occur, reaching $45 \%$ of the initial pure states at low temperature.
Further,  for every choice of temperature and pumping frequency, it is possible to find 
some initial state for which violations of the second law of thermodynamics occur repeatedly in time, becoming more and more typical for temperatures below $0.006$ $K$. 
Violations of the second law of thermodynamics are therefore not exceptional, rather they are inherent to the non complete positivity of the considered Redfield dynamics.
Whether such violations of the second law of thermodynamics are a feature of all non-completely positive dissipative dynamics is an open question; an answer to it  would demand 
either the proof that complete positivity is not only sufficient but also necessary to the non-negativity of the internal entropy production or devising an example of non-completely positive 
dissipative dynamics that does not conflict with thermodynamic expectations.
Both tasks would require a stronger characterisation of the generators of positive, but not completely positive dynamical maps, an issue which is still an open problem both mathematically and physically. 
\end{remark}

\subsection{Currents}

From an experimental point of view, due to the high time-resolution achieved by the present measurement devices, discriminating the behavior of the internal entropy production in the Redfield and completely positive cases is in line of principle perfectly possible through a tomographic reconstruction of the time-evolving state. However, a more direct check of complete positivity in this model is possible by studying the asymptotic current supported by the micro-circuit. Indeed, the model was devised as to provide an experimental setting for probing the fate of electronic currents in open quantum micro-circuits.
These specific characteristics can now be used to probe the complete positivity of the micro-circuit reduced dynamics; in fact, by numerical integration of the master equations made with the parameters chosen in~\cite{tosatti}, the temporal patterns of the currents supported by the micro-circuit under the Redfield and completely positive dynamics appear different enough to allow for an experimental test.
This fact may enable one to discriminate which one of the two possible dynamics is actually the best description for the true dynamics, thereby sorting out the fate of the second law of thermodynamics in the present model.

\begin{figure}[ht]
\centering
\includegraphics[width=0.50\textwidth]{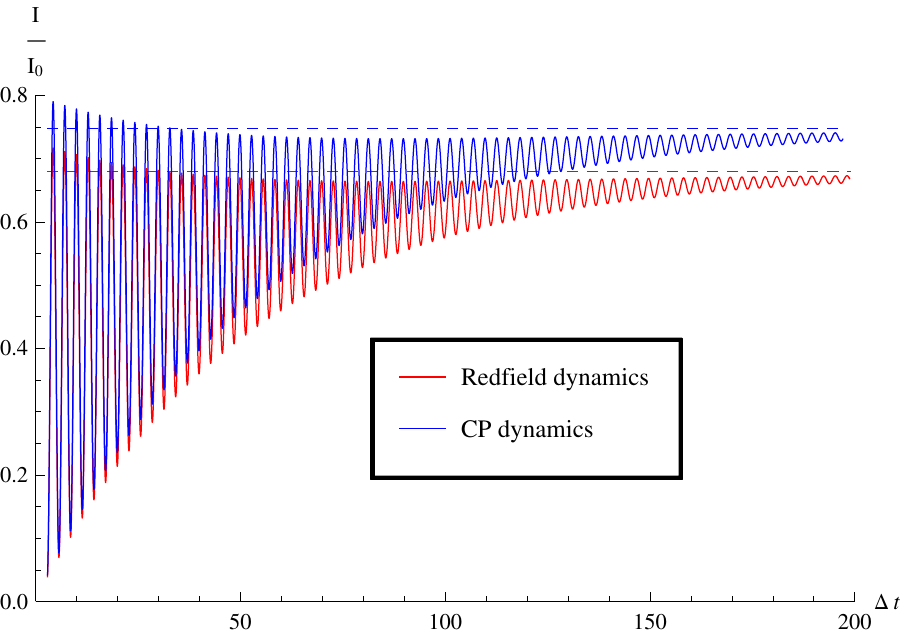}
\caption{\small Time behaviour of the current supported by the micro-circuit; dashed lines represent the asymptotic values.}
\label{fig:curr1}
\end{figure}

\begin{figure}[ht]
\centering
\includegraphics[width=0.50\textwidth]{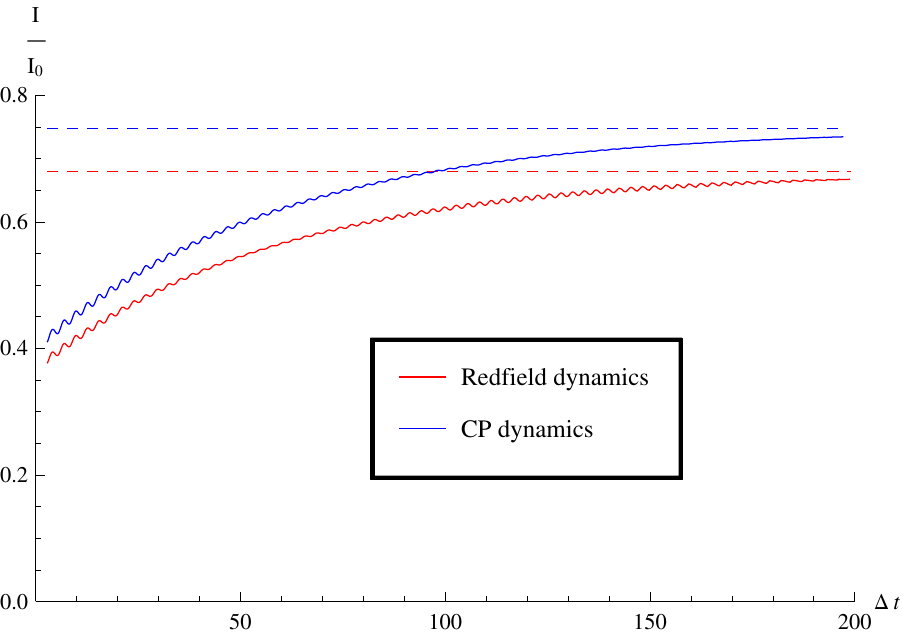}
\caption{\small Time behaviour of the current supported by the micro-circuit mediated over two periods; dashed lines represent the asymptotic values.}
\label{fig:curr2}
\end{figure}

The first of the two following figures shows the differences in the time-behaviour of the current supported by the two reduced dynamics starting with the initial state considered in Figure~\ref{fig:sigma_100_2} above. The second one refers to the fact that, despite the extremely high time-resolutions  nowadays achievable experimentally, what one may hope to
observe are not the pure oscillations in Figure~\ref{fig:curr1}, but oscillations mediated over a few periods.
Despite of this, as shown in Figure~\ref{fig:curr2}, it looks possible to discriminate between the completely and non-completely positive patterns.

\vfill\break

The possibility of discriminating between the two types of dynamics does not only hold for the specific values of the physical parameters considered above as chosen in~\cite{tosatti}; indeed, as shown in the following Figures~\ref{fig6} and~\ref{fig7}, for temperatures $T\leq 0.1\, K$ and rates $\Omega/\Delta$ in the range $0.3-10$, the stationary states for the Redfield and completely positive dynamics support current behaviours different enough to be amenable to experimental tests.

\begin{figure}[H]
\centering
\includegraphics[width=0.50\textwidth]{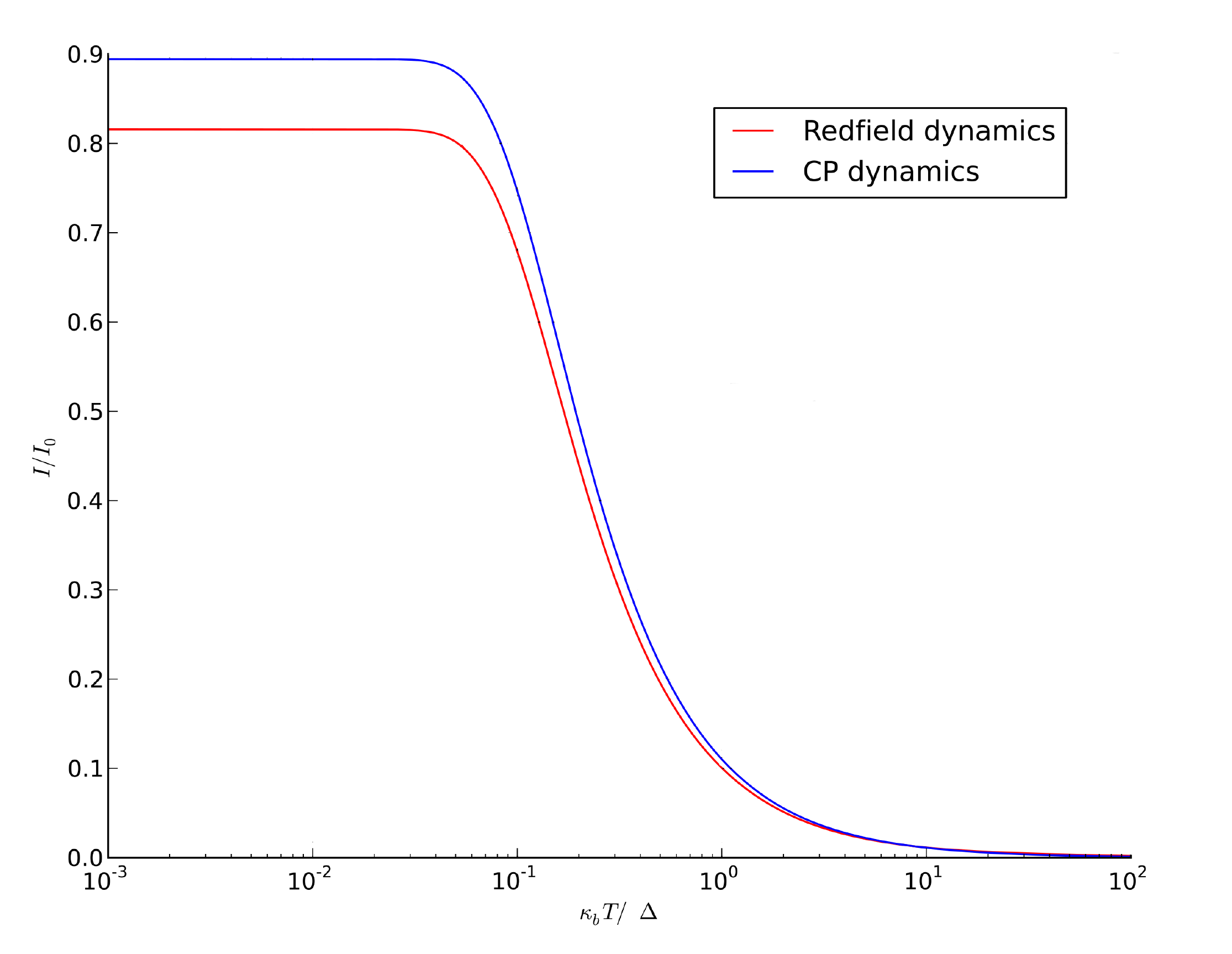}
\caption{\small Asymptotic current as function of the temperature, with $\lambda=0.005$ and $\Omega=2\Delta$ and initial state as in Figure~\ref{fig:sigma_100_2}.}
\label{fig6}
\end{figure}

\begin{figure}[H]
\centering
\includegraphics[width=0.50\textwidth]{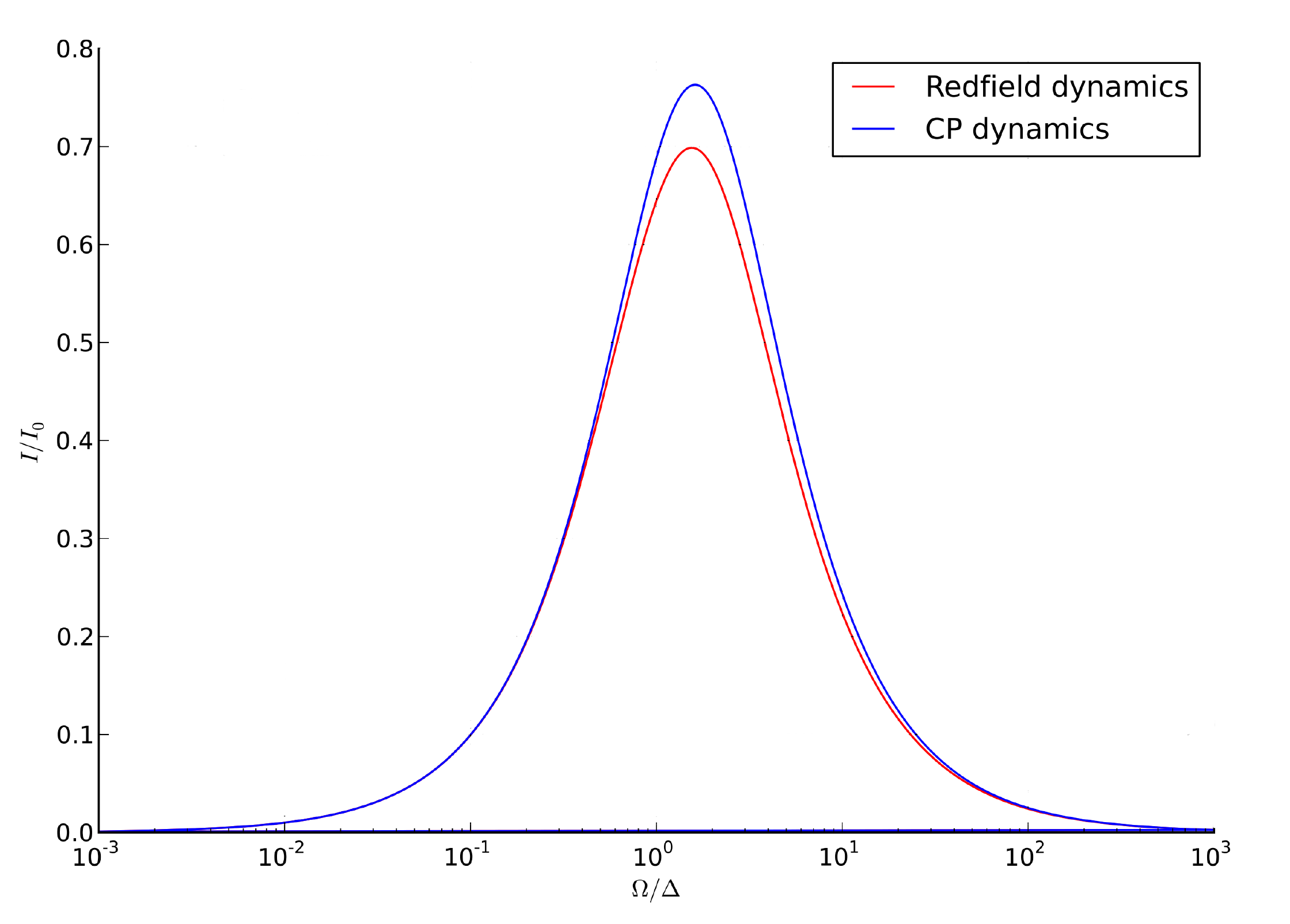}
\caption{\small Asymptotic current as function of $\Omega/\Delta$ with $\lambda=0.005$, temperature such that $\kappa T/\Delta=0.1$ and initial state as in Figure~\ref{fig:sigma_100_2}.}
\label{fig7}
\end{figure}

\section{Conclusions}

A typical argument against the request of complete positivity is that what really physically matters in the case of an open quantum system dynamics is the preservation of the positivity of the eigenvalues of time-evolving density matrices, for they must be interpreted as probabilities.
From this point of view, advocating the possible entanglement of the system of interest with an auxiliary inert system in order to justify the request of complete positivity looks like an abstract constraint.

We have here showed that complete positivity cannot so easily be dismissed because of its connection with the second law of thermodynamics. Indeed, although it was known that complete positivity is sufficient to achieve non-negative internal entropy production, no concrete, experimentally appealing models have been proposed to check the connection between the structure of an open quantum dynamics and possible violations of the second law of thermodynamics.
We have thus considered a model of open driven quantum micro-circuit which, in line of principle, can be experimentally studied, and showed that, if its dissipative dynamics is described by a Redfield dynamics, it would conflict with the principle of non-negative internal entropy production on a large variety of initial states, either at time $t=0$ or repeatedly in the course of time. This latter fact is particularly significant as, in general, unphysical effects due to lack of complete positivity are thought to occur on very short time-scales (a similar situation in the case of non-Markovian effects is discussed in \cite{breuer12}).  

The model was originally proposed to experimentally probe the electronic current sustained by the micro-circuit in weak interaction with its thermal environment: we have shown that looking at the temporal behaviour of the current may provide an experimental way to discriminate between the non-completely positive Redfield dynamics used in the model and the completely positive one that we have derived by adapting to a time-dependent Hamiltonian the standard weak-coupling limits techniques.
In turn, this would provide a check of the validity of the second law of thermodynamics at the quantum microscopic level.

Indeed, the violations of the second law are not related to exceptional initial states or values of the physical parameters of the model, rather they are typical and related to the structure of the Redfield dynamics which we have shown to be not even positivity-preserving.
Whether the conflict with thermodynamics occur for general non-completely positive dynamics is an open question: it is not 
known that complete positivity is also necessary to yield non-negative entropy production, nor there exist examples of only positivity-preserving dynamics that do not violate the second law of thermodynamics.
Indeed, on one hand, unlike for completely positive time-evolutions, there is no general characterisation of the generators of only positive ones; on the other hand, a microscopic derivation of a non-completely positive dissipative dynamics is very likely to yield a not even positivity-preserving time-evolution.
Hopefully, the findings of this manuscript will stimulate further research along these lines.
\bigskip

\noindent
\textbf{Acknowledgments}\quad
M.P. gratefully acknowledges the support from Funda\c{c}\~{a}o para a  
Ci\^{e}ncia e a Tecnologia (Portugal), namely through programmes  
PTDC/POPH and projects PEst-OE/EGE/UI0491/2013,  
PEst-OE/EEI/LA0008/2013, IT/QuSim and\\ 
CRUP-CPU/CQVibes, partially  
funded by EU FEDER, from the EU FP7 project LANDAUER (GA 318287), and  
support from FCT through scholarship SFRH/BD/52240/2013.

\appendix

\section{Redfield type master equation}
\label{appA}

In this appendix we discuss the derivation of the master equation~\eqref{eq:metosatti}: we follow the so-called projection technique~\cite{N58,Z60} applied
to the interaction representation.
We thus set
\bea
\label{appint1aa}
R_{\cs\cb}(t)&=&{\rm e}^{it(H_\eff+H_\cb)}\, {\varrho}_{\cs\cb}(t)\,{\rm e}^{-it(H_\eff+H_\cb)}\\
\label{appintab}
K_{\cs\cb}(t)&=&{\rm e}^{it(H_\eff+H_\cb)}\, {H}_{\cs\cb}(t)\,{\rm e}^{-it(H_\eff+H_\cb)}\ ,
\eea
whence
\be
\label{appint1b}
\frac{{\rm d}R_{\cs\cb}(t)}{{\rm d}t}=\lambda\,\KK_t[R_{\cs\cb}(t)]\ ,\quad \KK_t[R_{\cs\cb}(t)]=-i\,[K_{\cs\cb}(t)\,,\,R_{\cs\cb}(t)]\ .
\ee
Let $\varrho_\beta$ be the bath Gibbs state at temperature $T$ such that $[H_\cb\,,\,\varrho_\beta]=0$;
the following linear operators act as projectors on the states $ {\varrho}_{\cs\cb}(t)$ of the compound system $\cs+\cb$:
\be
\label{appint1c}
\PP[R_{\cs\cb}(t)]=\Big(\tr_\cb\Big(R_{\cs\cb}(t)\Big)\Big)\otimes \varrho_\beta=R(t)\otimes\varrho_\beta\ ,\quad \QQ=\id -\PP\ ,
\ee
with
\be
\label{appint1d}
R(t)=\tr_\cb \Big(R_{\cs\cb}(t)\Big)={\rm e}^{itH_\eff}\,\tr_\cb\Big(\varrho_{\cs\cb}(t)\Big)\,{\rm e}^{-itH_\eff}
\ee
giving the time-evolving density matrix of the open quantum system $\cs$ in its own interaction representation.

Let $\KK^{PP}_t=\PP\circ\KK_t\circ\PP$, $\KK^{PQ}_t=\PP\circ\KK_t\circ\QQ$, $\KK^{QP}_t=\QQ\circ\KK_t\circ\PP$ and $\KK^{QQ}_t=\QQ\circ\KK_t\circ\QQ$ where $\circ$ denotes the composition of maps. Then,~\eqref{appint1b} splits into the two coupled differential equations
\bea
\label{appint2a}
\frac{{\rm d}\PP[R_{\cs\cb}(t)]}{{\rm d}t}&=&\lambda\,\KK^{PP}_t\circ\PP[R_{\cs\cb}(t)]+\lambda\,\KK^{PQ}_t\circ\QQ[R_{\cs\cb}(t)]\\
\label{appint2b}
\frac{{\rm d}\QQ[R_{\cs\cb}(t)]}{{\rm d}t}&=&\lambda\,\KK^{QP}_t\circ\PP[R_{\cs\cb}(t)]+\lambda\,\KK^{QQ}_t\circ\QQ[R_{\cs\cb}(t)]\ .
\eea
The second equation is formally solved by
\be
\label{appint3a}
\QQ[R_{\cs\cb}(t)]=\WW_{t,0}^{QQ}\circ\QQ[R_{\cs\cb}]+\lambda\,\int_0^t{\rm d}s\,
\WW_{t,s}^{QQ}\circ\KK^{QP}_s\circ\PP[R_{\cs\cb}(s)]\\
\ee
with $\WW_{t,s}^{QQ}$ the time-ordered solution to
\be
\label{appint3b}
\frac{{\rm d}\WW_{t,s}^{QQ}}{{\rm d}t}=\lambda\,\KK^{QQ}_t\circ\WW_{t,s}^{QQ}\ , \qquad \WW_{s,s}={\rm id}\ .
\ee
With initial condition $ {\varrho}_{\cs\cb}=\varrho\otimes\varrho_\cb$, from
$\QQ[R_{\cs\cb}]=\QQ[ {\varrho}_{\cs\cb}]=0$ one gets
\be
\label{appint3c}
\QQ[R_{\cs\cb}(t)]=\lambda\,\int_0^t{\rm d}s\,
\WW_{t,s}^{QQ}\circ\KK^{QP}_s\circ\PP[R_{\cs\cb}(s)]\ ,
\ee
whence, inserting~\eqref{appint3c} into~\eqref{appint2a},
\be
\label{appint4a}
\frac{{\rm d}\PP[R_{\cs\cb}(t)]}{{\rm d}t}=\lambda\,\KK^{PP}_t\circ\PP[R_{\cs\cb}(t)]+\lambda^2\,\int_0^t{\rm d}s\,
\KK^{PQ}_t\circ\WW^{QQ}_{t,s}\circ\KK_s^{QP}\circ\PP[R_{\cs\cb}(s)]\ .
\ee
The form of the interaction Hamiltonian in equation~\eqref{ME1b} and the fact that position operators have vanishing mean values with respect to Gibbs states yield $\tr_\cb\Big(\varrho_\cb K_{\cs\cb}(t)\Big)=0$; then, $\PP[R_{\cs\cb}(t)]=R(t)\otimes\varrho_\beta$ implies
\be
\label{appint4b}
\frac{{\rm d}R(t)}{{\rm d}t}=-\,\lambda^2
\int_0^t{\rm d}u\,\tr_\cb\Big(\Big[K_{\cs\cb}(t),\QQ\circ\WW^{QQ}_{t,u}\circ\QQ\Big[K_{\cs\cb}(u)\,,\,R(u)\otimes\varrho_\beta\Big]\Big]\Big)\ .
\ee
The above equation depends on the history of the system state $R(s)$ for all times $0\leq s\leq t$; in order to eliminate this dependence, one takes into account the weak-coupling hypothesis $\lambda\ll 1$ and looks at the dynamics as a function of a slow time parameter $\tau=t\lambda^2$.
Firstly, by a change of integration variable $s=t-u$, \eqref{appint4b} is recast as
\be
\label{appint4c}
\frac{{\rm d}R(t)}{{\rm d}t}=-\,\lambda^2
\int_0^t{\rm d}u\,\tr_\cb\Big(\Big[K_{\cs\cb}(t),\QQ\circ\WW^{QQ}_{t,t-u}\circ\QQ\Big[K_{\cs\cb}(t-u)\,,\,R_{t-u}\otimes\varrho_\beta\Big]\Big]\Big)\ .
\ee
Then, letting $\lambda\to 0$, $\WW^{QQ}_{t,s}\to{\rm id}$ as the right hand side of~\eqref{appint3b} vanishes, and
$$
\QQ\circ\WW^{QQ}_{t,s}\circ\QQ\Bigg[\Big[K_{\cs\cb}(s)\,,\,R(s)\otimes\varrho_\beta\Big]\Bigg]\,\to\,
\QQ\Bigg[\Big[K_{\cs\cb}(s)\,,\, R(s)\otimes\varrho_\beta\Big]\Bigg]=\Big[K_{\cs\cb}(s)\,,\, R(s)\otimes\varrho_\beta\Big]\ .
$$
The last equality follows from  $\tr_\cb\Big(\varrho_\beta K_{\cs\cb}(s)\Big)=0$, as explained before.

At this point, one usually sends the integration upper limit to $+\infty$ and, in $R(t-u)$, replaces $t-u=\tau/\lambda^2-u$ with $t$;  then,
the equation~\eqref{appint4c} reads
\be
\label{appint6a}
\frac{{\rm d}R(t)}{{\rm d}t}=-\lambda^2
\int_0^{+\infty}{\rm d}u\,\tr_\cb\Big(\Big[K_{\cs\cb}(t)\,,\,\Big[K_{\cs\cb}(t-u)\,,\,R(t)\otimes\varrho_\beta\Big]\Big]\Big)\ .
\ee
By going back from the interaction picture to the Schr\"odinger one, the following master equation for $\varrho(t)$ is finally obtained,
\bea
\label{appint6b}
\frac{{\rm d}\varrho(t)}{{\rm d}t}&=&\LL_t[\varrho(t)]=-i\,\Big[H_\eff\,,\, \varrho(t)\Big]+\lambda^2\, {\NN}_t[\varrho(t)]\\
 {\NN}_t[\varrho(t)]&=&-
\int_0^{+\infty}{\rm d}u\,\tr_\cb\Big(\Big[ {H}_{\cs\cb}(t)\,,\,
\Big[{\rm e}^{u(\HH_\eff+\HH_\cb)}[ {H}_{\cs\cb}(t-u)]\,,\, \varrho(t)\otimes\varrho_\beta\Big]\Big]\Big)\ ,
\label{appint6c}
\eea
where
\be
\label{aid1}
{\rm e}^{u(\HH_\eff+\HH_\cb)}[X]={\rm e}^{-iu(H_\eff+H_\cb)}\,X\,{\rm e}^{iu(H_\eff+H_\cb)}\ .
\ee
Using~\eqref{ME1b} and the thermal state $2$-point functions~\eqref{2pointf} one finally obtains:
\bea
\nonumber
 {\NN}_t[ \varrho(t)]&=&-\sum_{\xi=1,3}\sum_n\lambda^2_n
\int_0^{+\infty}{\rm d}u\,\Big\{C(\omega_n,u)\Big[ {\sigma}_\xi(t)\,,\,
{\rm e}^{u\HH_\eff}[ {\sigma}_\xi(t-u)]\, \varrho(t)\Big]\\
\label{appint7a}
&+&C^*(\omega_n,u)\Big[ \varrho(t)\,{\rm e}^{u\HH_\eff}[ {\sigma}_\xi(t-u)]\,,\, {\sigma}_\xi(t)\Big]\Big\}\ ,
\eea
where
\be
\label{aid0}
C(\omega_n,u)=2m\omega_n\tr_\cb\Big(\varrho_\beta\,q_{\xi,n}\,{\rm e}^{-iuH_\cb}\,q_{\xi,n}\,
{\rm e}^{iuH_\cb}\Big)\ .
\ee
Since the couplings $\lambda_n$ do not depend on $\xi=1,3$  the explicit dependence on time $t$ disappears.
Indeed, let
\be
\label{appint8a}
\calr(t)=\begin{pmatrix}
\cos\Omega t&0&\sin\Omega t\\
0&1&0\\
-\sin\Omega t&0&\cos\Omega t
\end{pmatrix}
\ee
be the matrix which implements the rotation~\eqref{ME1b}:
\be
\label{appint8b}
 {\sigma}_\xi(t)=\sum_{\eta=1,2,3}\calr_{\xi\eta}(t)\sigma_\eta\ ,\quad \xi=1,3\ .
\ee
Then, for generic $2\times 2$ matrices $A$ and $B$ one finds
\bean
\sum_{\xi=1,3}A\, {\sigma}_\xi(t)\,B\, {\sigma}_\xi(t-u)&=&
\sum_{\xi=1,3}\sum_{\eta_1,\eta_2=1,2,3}\calr_{\xi\eta_1}(t)\calr_{\xi\eta_2}(t-u)\,A\, {\sigma}_{\eta_1}\,B\, {\sigma}_{\eta_2}\\
&=&\sum_{\eta_1,\eta_2=1,3}\calr_{\eta_1\eta_2}(-u)\,A\, {\sigma}_{\eta_1}\,B\, {\sigma}_{\eta_2}=
\sum_{\xi=1,3}A\,\sigma_\xi\,B\, {\sigma}_\xi(-u)\ .
\eean
Therefore, $ {\NN}_t[ \varrho(t)]$ becomes time-independent and equals
\bea
\nonumber
 {\NN}[ \varrho(t)]
&=&-\sum_{\xi=1,3}\sum_n\lambda^2_n
\int_0^{+\infty}{\rm d}u\,\Big\{C(\omega_n,u)\Big[\sigma_\xi\,,\,
{\rm e}^{-iu H_\eff}\, {\sigma}_\xi(-u)\,{\rm e}^{iu H_\eff} \varrho(t)\Big]\\
\label{appint8c}
&+&C^*(\omega_n,u)\Big[ \varrho(t)\,{\rm e}^{-iu H_\eff}\, {\sigma}_\xi(-u)\,{\rm e}^{iu H_\eff}\,,\,\sigma_\xi\Big]\Big\}\ ,
\eea
and, using~\eqref{eq:eqcorr} and~\eqref{spdens}, the master equation~\eqref{appint6b} reduces to~\eqref{eq:metosatti}.

Therefore, the generator $\LL_t$ in~\eqref{appint6b} becomes time-independent, too: $\LL_t=\LL$.
In order to recast it in Lindblad form as in~\eqref{eq:thermme} and~\eqref{offdiagdiss}, we first
pass from the Pauli triple $\sigma_{1,2,3}$ to the rotated one, $\hat{\sigma}_{1,2,3}$, in~\eqref{eq:newsigma}:
\be
\label{rot1a}
\sigma_\xi=\sum_{j=1}^3\mathcal{V}_{j\xi}\hat{\sigma}_j\ ,\qquad \cv=\frac{1}{\omef}\begin{pmatrix}\omega_\eff&0&0\\0&\Delta&\Omega\\0&-\Omega&\Delta\end{pmatrix}\ .
\ee
Then, $\displaystyle H_\eff=\frac{\omef}{2}\,\hat{\sigma}_3$ and
\be
\label{rot1b}
{\rm e}^{itH_\eff}\hat{\sigma}_j{\rm e}^{-itH_\eff}=\sum_{j,k=1}^3\cu^{jk}_\eff(t)\hat{\sigma}_k\ ,\quad
\cu_{\eff}(t)=\begin{pmatrix}\cos(\omef t)&-\sin(\omef t)&0\\\sin(\omef t)&\cos(\omef t)&0\\
0&0&1\end{pmatrix}\ .
\ee
One can thus rewrite the term $ {\NN}[ \varrho(t)]$ as follows:
\be
\label{appin9a}
{\NN}[ \varrho(t)]
=-\sum_{j,k=1}^3\int_0^{+\infty}{\rm d}u\,\cz_{jk}(-u)\,\Big\{G(u)\,\Big[\hat{\sigma}_j\,,\,
\hat{\sigma}_k\, \varrho(t)\Big]
+G^*(u)\,\Big[ \varrho(t)\,\hat{\sigma}_{k}\,,\,\hat{\sigma}_j\Big]\Big\}\ ,
\ee
where, $G(u)$ is as in~\eqref{eq:eqcorr}. By taking into account that $\xi=1,3$ the coefficients $\cz_{jk}(t)$ can be regrouped into the following matrix
by introducing the projection $\cp=\text{diag}(1,0,1)$:
\be
\label{appint9b}
\cz(t)=\cv\cp\calr(t)\cv^T\cu_\eff(t)=\begin{pmatrix}
cC+\frac{\Omega}{\omef}sS
&
\frac{\Omega}{\omef}cS-sC
&
\frac{\Delta}{\omef}S\\
\\
\frac{\Omega^2}{\omef^2}sC-\frac{\Omega}{\omef}cS
&
\frac{\Omega^2}{\omef^2}cC+\frac{\Omega}{\omef}sS
&
\frac{\Omega\Delta}{\omef^2}C\\
\\
\frac{\Omega\Delta}{\omef^2}sC-\frac{\Omega}{\omef}cS
&
\frac{\Omega\Delta}{\omef^2}cC+\frac{\Delta}{\omef}sS
&
\frac{\Delta^2}{\omef^2}C
\end{pmatrix}\ ,
\ee
where $c=\cos(\omef t)$, $s=\sin(\omef t)$ and $C=\cos(\Omega t)$, $S=\sin(\Omega t)$.

Finally, by separating the purely dissipative contribution $\DD[\varrho(t)]$ to $ {\NN}[ \varrho(t)]$ from the one corresponding to a Lamb-shift Hamiltonian
$H_{LS}$, one gets the right hand side of~\eqref{appint6b} as follows:
\bea
\label{appint10a}
{\LL}[ \varrho(t)]&=&-i\,\Big[H_\eff+\lambda^2\,{H}_{LS}\,,\, \varrho(t)\Big]+\lambda^2\,\DD[\varrho(t)]\\
\label{appint10aa}
\DD[\varrho(t)]&=&\sum_{j,k=1}^3{K}_{jk}\Big(\hat{\sigma}_k\, \varrho(t)\,\hat{\sigma}_j-\frac{1}{2}\Big\{\hat{\sigma}_j\hat{\sigma}_k\,,\,
 \varrho(t)\Big\}\Big)\\
\label{appint10b}
 {K}_{jk}&=&\int_0^{+\infty}{\rm d}u\,\Big(G(\tau)\,\cz_{jk}(-u)\,+\,G^*(\tau)\,\cz_{kj}(-u)\Big)= {K}^*_{kj}\ ,
\eea
where $\displaystyle {H}_{LS}=\sum_{j,k=1}^3 {H}_{jk}\,\hat{\sigma}_j\hat{\sigma}_k$, with
\be
\label{appint10c}
 {H}_{jk}=\frac{1}{2i}\int_0^{+\infty}{\rm d}\tau\,\Big(G(\tau)\,\cz_{jk}(-u)\,-\,G^*(\tau)\,\cz_{kj}(-u)\Big)= {H}^*_{kj}\ .
\ee
In order to recast the action of $ {\LL}[ \varrho(t)]$ as that of a $4\times 4$ matrix $-2\cl$ on the Bloch vector $(1, {r}_1(t), {r}_2(t), {r}_3(t))$ as in~\eqref{Bloch}, one considers the linear action of $ {\LL}$ on the Pauli matrices $\hat{\sigma}_{1,2,3}$ and on the identity $\hat{\sigma}_0=1$:
$ {\LL}[\hat{\sigma}_\mu]=\sum_{j=1}^3 {\cl}_{j\mu}\hat{\sigma}_j$.
With $ {r}_0(t)=1$ because of trace conservation, this gives
\bea
\label{appint11a}
 {\LL}[ \varrho(t)]&=&\frac{1}{2}\Big( {\LL}[1]+\sum_{j=1}^3 {r}_j(t) {\LL}[\hat{\sigma}_j]\Big)=
-\sum_{j=1}^3\Big(\sum_{\mu=0}^3 {\cl}_{j\mu} {r}_\mu(t)\Big)\,\hat{\sigma}_j\\
\label{appint11b}
 {\cl}&=&{\ch}_\eff\,+\lambda^2\, {\ch}_{LS}\,+\lambda^2\, {\cd}=\begin{pmatrix}0&0&0&0\\
 {\cl}_{10}& {\cl}_{10}& {\cl}_{10}& {\cl}_{10}\\
 {\cl}_{20}& {\cl}_{21}& {\cl}_{22}& {\cl}_{23}\\
 {\cl}_{30}& {\cl}_{31}& {\cl}_{32}& {\cl}_{33}
\end{pmatrix}\ .
\eea
The matrix $ {\cl}$ consists of an antisymmetric Hamiltonian contribution $ {\ch}_\eff+\lambda^2\, {\ch}_{LS}$,
where
\bea
\label{appint11b1}
 {\ch}_\eff&=&\begin{pmatrix}0&0&0&0\\
0&0&\omef/2&0\\
0&-\omef/2&0&0\\
0&0&-0&0
\end{pmatrix}\\
\label{appint11c}
 {\ch}_{LS}&=&\begin{pmatrix}0&0&0&0\\
0&0& {\ch}_{12}& {\ch}_{13}\\
0&- {\ch}_{12}&0& {\ch}_{23}\\
0&- {\ch}_{13}&- {\ch}_{23}&0
\end{pmatrix}\ ,\quad
\left\{\begin{matrix}
 {\ch}_{12}=&2\,\ci m( {H}_{21})\\
 {\ch}_{13}=&2\,\ci m( {H}_{31})\\
 {\ch}_{23}=&2\,\ci m( {H}_{32})
\end{matrix}
\right.\ ,
\eea
plus a purely dissipative term
\bea
\label{appint11e}
 {\cd}&=&\begin{pmatrix}0&0&0&0\\
 {\ck}_{10}& {\ck}_{11}& {\ck}_{12}& {\ck}_{13}\\
 {\ck}_{20}& {\ck}_{12}& {\ck}_{22}& {\ck}_{23}\\
 {\ck}_{30}& {\ck}_{13}& {\ck}_{23}& {\ck}_{33}\\
\end{pmatrix}\ ,\
\left\{
\begin{matrix}
 {\ck}_{10}&=\ci m( {K}_{23})\\
 {\ck}_{20}&=\ci m( {K}_{31})\\
 {\ck}_{30}&=\ci m( {K}_{12})
\end{matrix}\right.\\
\label{appint11f}
&&
\left\{\begin{matrix}
 {\ck}_{11}&= {K}_{22}+ {K}_{33}\\
 {\ck}_{22}&= {K}_{11}+ {K}_{33}\\
 {\ck}_{33}&= {K}_{11}+ {K}_{22}
\end{matrix}\right.\ ,\qquad
\left\{\begin{matrix}
 {\ck}_{12}&=-\calr e( {K}_{12})\\
 {\ck}_{13}&=-\calr e( {K}_{13})\\
 {\ck}_{23}&=-\calr e( {K}_{23})
\end{matrix}\right.\ .
\eea
Using the expressions in~\eqref{appint9b},~\eqref{appint10b} and~\eqref{appint10c}, the matrix entries explicitly read
\bea
\nonumber
 {\ck}_{10}&=&\frac{\Delta}{\omef}\int_0^{+\infty}{\rm d}u\int_0^{+\infty}{\rm d}\omega\,J(\omega)\,\sin(\omega u)\,\Big(
\sin(\omef u)\sin(\Omega u)+\\
\label{appint12a}
&&
\hskip 3cm+\frac{\Omega}{\omef}\Big(\cos(\omef u)-1\Big)\cos(\Omega u)
\Big)
\\
\nonumber
 {\ck}_{20}&=&\frac{\Delta}{\omef}\int_0^{+\infty}{\rm d}u\int_0^{+\infty}{\rm d}\omega\,J(\omega)\,\sin(\omega u)\,\Big(
-\Big(1+\cos(\omef u)1\Big)\sin(\Omega u)\Big)+\\
\label{appint12b}
&&
\hskip 3cm+\frac{\Omega}{\omef}\sin(\omef u)\cos(\Omega u)
\Big)
\\
\nonumber
 {\ck}_{30}&=&\int_0^{+\infty}{\rm d}u\int_0^{+\infty}{\rm d}\omega\,J(\omega)\,\sin(\omega u)\,\Big(
-\frac{2\Omega^2+\Delta^2}{\omef^2}\sin(\omef u)\cos(\Omega u)+\\
\label{appint12c}
&&
\hskip 3cm+2\frac{\Omega}{\omef}\cos(\omef u)\sin(\Omega u)\Big)
\eea
\bea
\nonumber
 {\ck}_{11}&=&2\int_0^{+\infty}{\rm d} u\int_0^{+\infty}{\rm d}\omega\,J(\omega)\,\cos(\omega u)\coth(\frac{\beta\omega}{2})\,\Big(
\frac{\Omega}{\omef}\sin(\omef u)\sin(\Omega u)+\\
\label{appint13a}
&&
\hskip 1cm+\frac{\Omega^2}{\omef^2}\cos(\omef u)\cos(\Omega u)+\frac{\Delta^2}{\omef^2}\cos(\Omega u)
\Big)
\\
\nonumber
\\
 {\ck}_{12}&=&-\frac{\Delta^2}{\omef^2}\int_0^{+\infty}{\rm d}u\int_0^{+\infty}{\rm d}\omega\,J(\omega)\,\cos(\omega u)\,
\coth(\frac{\beta\omega}{2})\,\sin(\omef u)\cos(\Omega u)
\label{appint13b}
\\
\nonumber
\\
\nonumber
 {\ck}_{13}&=&\frac{\Delta}{\omega_\eff}\int_0^{+\infty}{\rm d}u\int_0^{+\infty}{\rm d}\omega\,J(\omega)\,\cos(\omega u)\coth(\frac{\beta\omega}{2})\,\Big(
-\frac{\Omega}{\omef}\sin(\omef u)\cos(\Omega u)+\\
\label{appint13c}
&&
\hskip 1cm
+\sin(\Omega u)\Big(\cos(\omef u)-1\Big)\Big)
\eea
\bea
\nonumber
 {\ck}_{22}&=&2\int_0^{+\infty}{\rm d}u\int_0^{+\infty}{\rm d}\omega\,J(\omega)\,\cos(\omega u)\coth(\frac{\beta\omega}{2})\,\Big(
\cos(\omef u)\cos(\Omega u)+\\
\label{appint13d}
&&
\hskip 1cm+\frac{\Omega}{\omef}\sin(\omef u)\sin(\Omega u)+\frac{\Delta^2}{\omef^2}\cos(\Omega u)
\Big)\\
\\
\nonumber
\\
\nonumber
 {\ck}_{23}&=&-\frac{\Delta}{\omef}\int_0^{+\infty}{\rm d}u\int_0^{+\infty}{\rm d}\omega\,J(\omega)\,\cos(\omega u)\coth(\frac{\beta\omega}{2})\,\Big(
\sin(\omef u)\sin(\Omega u)+\\
\label{appint13e}
&&
\hskip 1cm+\frac{\Omega}{\omef}\Big(1+\cos(\omef u)\Big)\cos(\Omega u)\Big)
\\
\nonumber
\\
\nonumber
 {\ck}_{33}&=&2\int_0^{+\infty}{\rm d}u\int_0^{+\infty}{\rm d}\omega\,J(\omega)\,\cos(\omega u)\coth(\frac{\beta\omega}{2})\,\Big(
\frac{2\Omega^2+\Delta^2}{\omef^2}\cos(\omef u)\cos(\Omega u)+\\
&&\hskip 1cm +2\frac{\Omega}{\omef}\sin(\omef u)\sin(\Omega u)\Big)\ ;
\label{appint13f}
\eea
\bea
\nonumber
 {\ch}_{12}&=&\int_0^{+\infty}{\rm d} u\int_0^{+\infty}{\rm d}\omega\,J(\omega)\,\cos(\omega u)\coth(\frac{\beta\omega}{2})\,\Big(
\frac{2\Omega^2+\Delta^2}{\omef^2}\sin(\omef u)\cos(\Omega u)-\\
\label{appint14a}
&&
\hskip 1cm-2\frac{\Omega}{\omef}\cos(\omef u)\sin(\Omega u)\Big)
\\
\nonumber
\\
 {\ch}_{13}&=&\frac{\Delta}{\omef}\int_0^{+\infty}{\rm d}u\int_0^{+\infty}{\rm d}\omega\,J(\omega)\,\cos(\omega u)\,
\coth(\frac{\beta\omega}{2})\,\Big(\frac{\Omega}{\omef}\sin(\omef u)\cos(\Omega u)-\\
\label{appint14b}
&&
\hskip 1cm
-\sin(\Omega u)\Big(1+\cos(\omef u)\Big)\Big)
\\
\nonumber
\\
\nonumber
 {\ch}_{23}&=&\frac{\Delta}{\omef}\int_0^{+\infty}{\rm d}u\int_0^{+\infty}{\rm d}\omega\,J(\omega)\,\cos(\omega u)\coth(\frac{\beta\omega}{2})\,\Big(
-\sin(\omef u)\sin(\Omega\tau)+\\
\label{appint14c}
&&
\hskip 1cm+\frac{\Omega}{\omef}\Big(1-\cos(\omef u)\Big)\cos(\Omega u)\Big)\ .
\eea
In the main text, we have added a superscript ``$\Red$" to $\mathcal{L}$, $\mathcal{H}_{LS}$ and $\mathcal{D}$ in order to distinguish them from the analogous expressions pertaining to a completely positive dynamics which are obtained in the next Appendix.

\section{Completely positive master equation}
\label{appB}

A physically consistent reduced dynamics can be obtained by a more careful treatment; it leads to a completely positive time-evolution, thus avoiding all the inconsistencies of the Redfield dynamics used in \cite{tosatti}.
We again use the projection technique but we follow the analysis of~\cite{dumcke78} without passing to the interaction representation.
By repeating the arguments of the previous Appendix, one arrives at the following analog of equation~\eqref{appint4b}:
\bea
\nonumber
\frac{{\rm d} \varrho(t)}{{\rm d}t}&=&-i\,[H_\eff\,,\, \varrho(t)]\\
\label{app6a}
&-&\lambda^2
\int_0^t{\rm d}u\,\tr_\cb\Big(\Big[ {H}_{\cs\cb}(t)\,,\,\QQ\circ\UU^{QQ}_{t,s}\circ\QQ\Big[\Big[ {H}_{\cs\cb}(u)\,,\, \varrho(t)
\otimes\varrho_\cb\Big]\Big]\Big)\ .
\eea
As shown in \cite{dumcke78}, a sounder strategy than the one that led to equation~\eqref{appint6a} in the previous Appendix consists firstly in formally integrating~\eqref{app6a}, yielding
\bea
\nonumber
 \varrho(t)&=&{\rm e}^{t\HH_\eff}[ {\varrho}]\ -\\
\label{app7}
&&\hskip -1.5cm -\lambda^2
\int_0^t{\rm d}u\,\int_0^u{\rm d}v\,{\rm e}^{(t-u)\HH_\eff}\Bigg[
\tr_\cb\Big(\Big[ {H}_{\cs\cb}(u),\QQ\circ\UU^{QQ}_{u,v}\circ\QQ\Big[ {H}_{\cs\cb}(v)\,,\, {\varrho}(v)
\otimes\varrho_\cb\Big]\Big]\Big)\Bigg]\ .
\eea
Secondly, in changing the double integral into
\bean
&&
\int_0^t{\rm d}v\,\int_v^t{\rm d}u\,{\rm e}^{(t-u)\HH_\eff}\,\Bigg[\tr_\cb\Big(\Big[ {H}_{\cs\cb}(u),\QQ\circ\UU^{QQ}_{u,v}\circ\QQ\Big[ {H}_{\cs\cb}(v)\,,
\, \varrho(v)\otimes\varrho_\cb\Big]\Big]\Big)\Bigg]=\\
&&\hskip -1cm
=\int_0^t{\rm d}v\,\int_0^{t-v}{\rm d}w\,{\rm e}^{(t-v-w)\HH_\eff}\,\Bigg[
\tr_\cb\Big(\Big[ {H}_{\cs\cb}(v+w)\,,\,\QQ\circ\UU^{QQ}_{v+w,v}\,\circ\, \QQ\Big[ {H}_{\cs\cb}(v)\,,\, \varrho(v)
\otimes\varrho_\cb\Big]\Big]\Big)\Bigg]\ ,
\eean
and, finally, in going to the slow time-scale $\tau=t\lambda^2$, $\lambda<<1$ where, using~\eqref{aid1}, one replaces
$$
\QQ\circ\UU^{QQ}_{v+w,v}\circ\QQ\Big[\Big[ {H}_{\cs\cb}(v)\,,\, \varrho(v)
\otimes\varrho_\cb\Big]\Big]\quad \hbox{by}\quad
{\rm e}^{w(\HH_\eff+\HH_\cb)}\Bigg[\Big[ {H}_{\cs\cb}(v)\,,\, \varrho(v)
\otimes\varrho_\cb\Big]\Bigg]\ ,
$$
so that the second integral with respect to ${\rm d}w$ becomes
\be
\label{app7b}
\overline{\NN}_v[ \varrho(v)]=\int_0^{+\infty}{\rm d}w\,
{\rm e}^{-w\HH_\eff}\Bigg[\tr_\cb\Big(\Big[ {H}_{\cs\cb}(v+w)\,,\,
{\rm e}^{w(\HH_\eff+\HH_\cb)}\,\Big[\Big[ {H}_{\cs\cb}(v)\,,\, {\varrho}_{v}
\otimes\varrho_\cb\Big]\Big]\Big]\Big)\Bigg]\ .
\ee
This yields
$$
 \varrho(t)={\rm e}^{t\HH_\eff}[ {\varrho}]-\lambda^2\int_0^t{\rm d}v\,\overline{\NN}_v[ \varrho(v)]
$$
that solves the  master equation:
\bea
\label{app8a}
\frac{{\rm d} \varrho(t)}{{\rm d}t}&=&-i\,\Big[H_\eff\,,\, \varrho(t)\Big]+
\lambda^2\overline{\NN}_t[ \varrho(t)]\ .
\eea
By proceeding as in the previous Appendix, one recasts~\eqref{app8a} in the time-independent form
\bea
\label{app100a}
\frac{{\rm d} \varrho(t)}{{\rm d}t}&=&
\overline{\LL}[ \varrho(t)]=-i\,\Big[H_\eff\,,\, \varrho(t)\Big]+\lambda^2\,
\overline{\NN}[ \varrho(t)]\\
\label{app100b}
\overline{\NN}[ \varrho(t)]&=&-i\,\Big[\overline{H}_{LS}\,,\, \varrho(t)\Big]+\lambda^2\,\sum_{j,k=1}^3
 {K}_{jk}\Big(\hat{\sigma}_k\, \varrho(t)\,\hat{\sigma}_j-\frac{1}{2}\Big\{\hat{\sigma}_j\hat{\sigma}_k\,,\,
 \varrho(t)\Big\}\Big)\ ,
\eea
where $\overline{H}_{LS}=\sum_{j,k=1}^3 {H}_{jk}\hat{\sigma}_j\hat{\sigma}_k$ and, with respect to \eqref{appint10b} and \eqref{appint10c}, the coefficients now read
\bea
\label{app8c}
 {H}_{jk}&=&\frac{1}{2i}\int_0^{+\infty}{\rm d}u\,\Big(G(u)\,\cz_{kj}(u)\,-\,G^*(u)\,\cz_{jk}(u)\Big)= {H}^*_{kj}\\
\label{app8d}
 {K}_{jk}&=&\int_0^{+\infty}{\rm d}u\,\Big(G(u)\,\cz_{kj}(u)\,+\,G^*(u)\,\cz_{jk}(u)\Big)= {K}^*_{kj}\ .
\eea

\begin{remark}
\label{rem7}
In the Bloch representation, the generator $\overline{\LL}[ \varrho(t)]$ corresponds to the action on the Bloch vector $(1, {r}_1(t), {r}_2(t), {r}_3(t))$ of a $4\times 4$ matrix $\overline{\cl}= {\ch}_\eff\,+\lambda^2\,\overline{\ch}_{LS}\,+\lambda^2\,\overline{\cd}$, whose coefficients are all equal to the ones in~\eqref{appint12a}--~\eqref{appint14c} apart from $ {\ck}_{10}$, $ {\ck}_{12}$, $ {\ck}_{13}$  and $ {\ch}_{23}$ which have opposite signs.
\end{remark}
\medskip

The formal solutions to~\eqref{app100a} are given by
\be
\label{app9a}
\varrho(t)={\rm e }^{t\overline{\LL}}[\varrho]={\rm e}^{t\HH_\eff}[\varrho]+\lambda^2\int_0^t{\rm d}s\,{\rm e}^{t\HH_\eff}\circ\overline{\NN}[ {\varrho}_s]
\ee
with ${\rm e}^{t\HH_\eff}[\varrho]=\exp(-it\, H_\eff)\,\varrho\,\exp(it\, H_\eff)$.

On the slow time scale $\tau=\lambda^2\,t$, one rewrites
\be
\label{app9b}
{\rm e}^{-t\HH_\eff}\circ{\rm e}^{t\overline{\LL}}[\varrho]=\varrho+\int_0^\tau{\rm d}u\,
\Big\{{\rm e}^{-u/\lambda^2\HH_\eff}\circ \overline{\NN}\circ
{\rm e}^{u/\lambda^2\HH_\eff}\Big\}\Big[{\rm e}^{-u\HH_\eff}\circ{\rm e}^{u\overline{\LL}}[\varrho]\Big]\ ;
\ee
when $\lambda\to0$ the fast oscillations in the term in curly brackets average to zero. This allows one to replace that term
by its ergodic average
\be
\label{app9c}
\NN=\lim_{T\to+\infty}\frac{1}{2T}\int_{-T}^T{\rm d}s\,{\rm }^{-s/\lambda^2\HH_\eff}\circ\overline{\NN}\circ
{\rm }^{s/\lambda^2\HH_\eff}\ ,
\ee
which fulfils $\NN\circ\HH_\eff=\HH_\eff\circ\NN$, so that the resulting master equation is
\be
\label{app10}
\frac{{\rm d} \varrho(t)}{{\rm d}t}= {\LL}[ \varrho(t)]=-i\,\Big[H_\eff+\lambda^2\, {H}_{LS}\,,\, \varrho(t)\Big]
+\lambda^2 {\DD}[ \varrho(t)]
\ee
In the Bloch representation the action of $\NN$ corresponds to that of the $4\times 4$ matrix
\be
\label{app11}
 {\cn}=\lim_{T\to+\infty}\frac{1}{2T}\int_0^T{\rm d}s\,\cu_\eff(-s)\,\overline{\cn}\,\cu_\eff(s)\ ,
\ee
with $\overline{\cn}$ the $4\times 4 $ matrix corresponding to $\overline{\NN}$  and $\cu_\eff(s)$ the $4\times 4$
matrix with $1,0,0,0$ in the first row and column and, in the rest, the $3\times 3$ matrix in~\eqref{rot1b}.
Then, the action of $ {\LL}$ can be represented by means of the $4\times 4$ matrix
$ {\cl}=\ch_\eff+\lambda^2\, {\ch}_{LS}+\lambda^2\, {\cd}$, where $\ch_\eff$ is as in~\eqref{appint11b1}, while
\be
\label{app11a}
 {\ch}_{LS}=\begin{pmatrix}0&0&0&0\\
0&0& {\ch}_{12}&0\\
0&- {\ch}_{12}&0&0\\
0&0&0&0
\end{pmatrix}
\ ,\qquad
 {\cd}=\begin{pmatrix}0&0&0&0\\
0& \mathcal{K}_{11}+ \mathcal{K}_{22}&0&0\\
0&0& \mathcal{K}_{11}+ \mathcal{K}_{22}&0\\
 \mathcal{K}_{30}&0&0& \mathcal{K}_{33}
\end{pmatrix}\ .
\ee

\end{document}